%% file: GIB_arxiv_nov.tex
\documentclass[3p,times]{elsarticle}

\usepackage{graphicx}
\usepackage{subfigure}
\include{macrosKarpou}

\usepackage{etoolbox}
\patchcmd{\emailauthor}{(#2)}{}{}{}
\patchcmd{\urlauthor}{(#2)}{}{}{}

\usepackage{lineno}

\modulolinenumbers[5]

\journal{arXiv}

\begin{document}

\begin{frontmatter}

\title{GENERALIZED INTERNAL BOUNDARIES (GIB)}


\author[mymainaddress]{Georgios K. Karpouzas\corref{mycorrespondingauthor}}

\cortext[mycorrespondingauthor]{Corresponding author}
\ead{g.karpouzas@engys.com}

\author[mymainaddress]{ Eugene De Villiers}

\address[mymainaddress]{ENGYS Ltd, \\Studio 20, Royal Victoria Patriotic, John Archer Way, SW18 3SX, London, UK}
\ead[url]{www.engys.com}

\begin{abstract}
Representing large scale motions and topological changes in the finite volume (FV) framework, while at the same time preserving the accuracy of the numerical solution, is difficult. In this paper, we present a robust, highly efficient method designed to achieve this capability. The proposed approach conceptually shares many of the characteristics of the cut-cell interface tracking method, but without the need for complex cell splitting/merging operations. The heart of the new technique is to align existing mesh facets with the geometry to be represented. We then modify the matrix contributions from these facets such that they are represented in an identical fashion to traditional boundary conditions. The collection of such faces is named a Generalised Internal Boundary (GIB). In order to introduce motion into the system, we rely on the classical ALE (Arbitrary Lagrangian-Eulerian) approach, but with the caveat that the non-time-dependent motion of elements instantaneously crossing the interface is handled separately from the time-dependent component. The new methodology is validated through comparison with: $a)$ a body fitted grid simulation of an oscillating two dimensional cylinder and $b)$ experimental results of a butterfly valve.
\end{abstract}

\begin{keyword}
Immersed Boundary, Finite-Volume method, Mesh motion, CFD, ALE
\end{keyword}

\end{frontmatter}


\section{Introduction}

There are many important fluid-flow problems that occur in concert with complex moving boundaries. These applications range from the simulation of a beating heart to the burning of a solid propellant inside a rocket engine. Despite the fact that deforming boundaries with mesh motion engines can be applied to a wide range of moving mesh applications, there is always a limit to how much deformation can be accommodated within a fixed grid structure. Further, topological changes in the geometry typically require very complex manipulations that are difficult to generalise. To handle larger deformations using this approach, remeshing and solution mapping is required, which is computational expensive and prone to error.

To overcome these difficulties, the first immersed boundary method (IB) was intoduced by C.Peskin \cite{Peskin1972}\cite{Peskin1977}\cite{Peskin2002}, who worked in the blood flow simulation of a human heart. In his method, the immersed boundaries are modelled by inserting additional forcing terms in the momentum equation. The force terms are applied via a Dirac function in the cells close to the interface, which constrains the flow to follow the immersed boundary.
Later, direct forcing methods were introduced by J.Mohd-Yusof \cite{mohdyusof1997}\cite{FADLUN200035}\cite{Majumdar2001}, who inserted a forcing term, which is determined by the difference between the interpolated velocities in the boundary points and the desired boundary velocities. 
Improved versions of discrete forcing, using cells from the solid region (ghost cells) to approximate the boundary conditions \cite{Tseng_ghostCell}, increased the accuracy of the immersed boundaries. However, none of the above IB methods is conservative, which introduces significant problems in the solution of the state equations. The approach also lacks generality in that all systems related to it have to be implemented specifically for every aspect of the model. Finally, the boundary conditions are depended on the interpolation, which is not straightforward on arbitrarily complex meshes. 

A more rigorous approach is the cut-cell method \cite{YE1999209}\cite{Hartmann2011}\cite{SCHNEIDERS2013786}, where new faces are inserted in the Cartesian background mesh, which cut the cells in order to fit the surface locally. This method is conservative and accurate. However, the insertion of new faces and as a result of cells, breaks the connectivity of the mesh. This can slow down the whole simulation times due to updates in the decomposition, fields, mesh and matrix solver structures. Moreover, the splitting and merging of the cells is not straightforward in complex cases.

Overset-grid method \cite{TANG2003567}\cite{overset1}\cite{overset2} is used in applications with large motions. This method uses one background mesh and one body-fitted mesh. The two meshes communicate using volumetric mesh to mesh interpolation. The advantage of this method is that it can have body-fitted layer elements close to the solid boundaries, as typically the overlapping area of the two meshes is located away from the wall boundaries. However, the search algorithms required for the interpolation can be computationally expensive, resulting in poor performance and scaling for complex and/or large cases. Overset-grid implementations also suffer from conservation issues, due to the volumetric interpolation coupling. Beyond the accuracy concerns that this introduces, unphysical oscillations or divergence are common with these systems. Another limitation of overset-grid methods is that, while rigid body motion is well represented, deforming boundaries are no more flexibly handled than via static meshes. Topological changes also introduce issues that have to be handled via methods similar to immersed boundaries.

In this paper we propose a novel technique which is similar to the cut-cell method, in that the interface is directly represented by the grid elements. However, the mesh connectivity (topology) remains unchanged. Instead of inserting new facets into the background mesh to represent the interface, we change the point coordinates near the interface such that existing facets are coincident with the interface location. The  GIB are constructed on the fly, using the collection of internal faces that are coincident with the boundary to be represented. The GIB facets now contribute to the matrix coefficients and source terms in an identical fashion to traditional boundaries. Therefore, the GIB boundaries behave just like normal external boundaries, but have the flexibility to occur at any position inside the original computational domain.
In order to support large transient deformations, it is necessary for cells to transition from one side of the interface to the other. Thus, we split the mesh fluxes into two parts. The first part consists of the flux due to the cell interface transition and is used to conservatively redistribute transported quantities to/from neighbouring cells. The remaining flux is the boundary flux and it is used in a classic ALE approach.

\section{Methodology}\label{Concept}

\begin{figure}[h!]
\centering
\subfigure[Interface and base mesh]
{
{\includegraphics[width=7.5cm,angle=0]{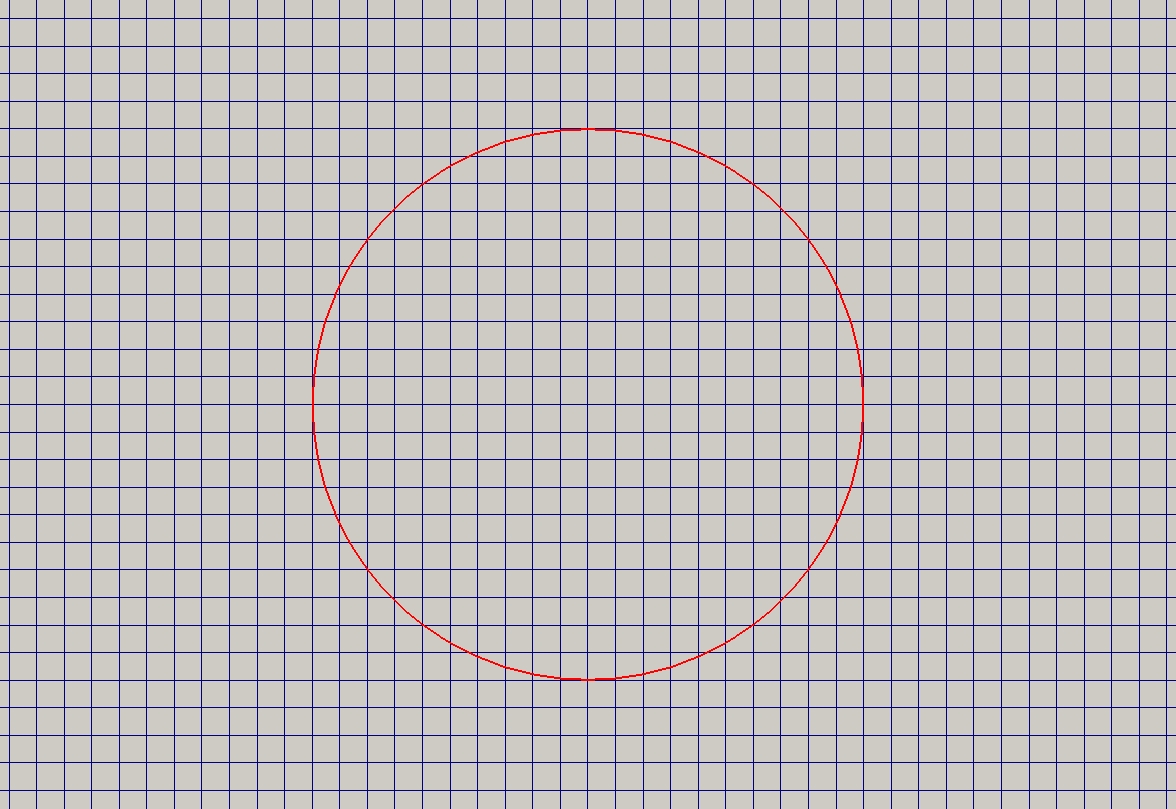}}
\label{gib_bef}
}
\subfigure[Locally conformed faces on the interface]
{
{\includegraphics[width=7.5cm,angle=0]{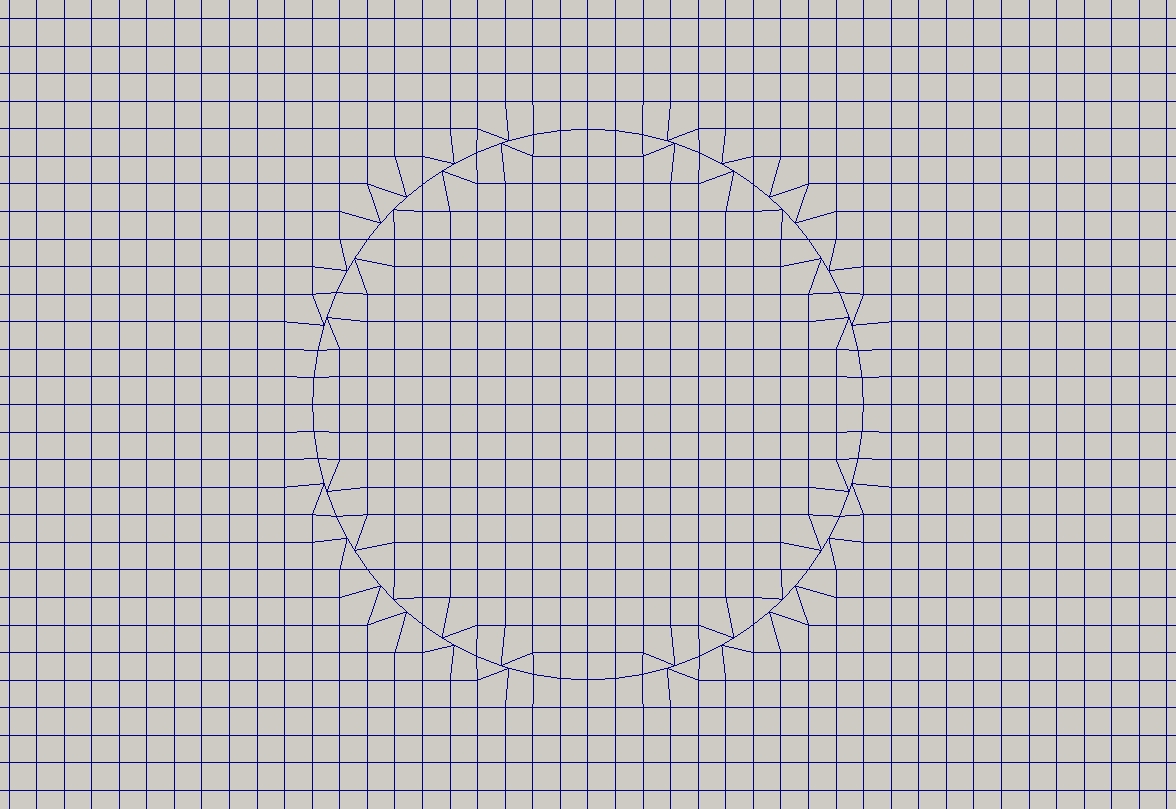}}
\label{gib_aft}
}
\caption{Solid/fluid interface (red line) with the background computational domain (left); Mesh after conforming intersecting elements to the surface (right)}
\label{gibConcept}
\end{figure}

The underlying idea behind the GIB is presented in figure \ref{gibConcept}, where the red line represents the solid/fluid surface and the grey area is the background computational domain (base mesh). In our implementation, the interface is represented using an external triangulated surface in STL format but it can be extended to every sharp interface method. The first step is to identify the edge sections of the base mesh graph that intersect the surface. These edge sections map directly to facets in the base mesh that bound each cell and through them the point locations that define the facets. The second step is to move these points such that the base mesh faces that map to an intersected graph edge become coincident with the surface. The collection of such faces constitutes the GIB and is made to behave in an indistinguishable manner from normal peripheral boundary elements.

In a face based implementation of the finite volume method \cite{Wesseling:2000:PCF:1211014}, the discretization of the governing equations results in a face specific matrix contribution (implicit or explicit) to the elements neighbouring each facet. These contributions depend on the type of the term (convection, diffusion, source etc), discretization schemes and algorithms \cite{ferziger2001}. Facets internal to the mesh typically contribute to two neighbour cells, while peripheral (boundary) faces contribute only to one neighbouring element. 

An example is given in equation \ref{gauss}, where the volume integral is transformed to a surface integral using Gauss's theorem. In the second order accurate discretized form, each face contributes to the gradient of all neighbouring cells.

\begin{eqnarray}   
		\int_{\Omega} \frac{\pp \xi}{\pp x_j} d\Omega
		= 
		\int_{S} \xi n_j dS
		=
		\sum_{i=1}^{n_f} \left({{S_f}_j} {\xi_f}\right)_i
		\label{gauss} 
\end{eqnarray}
where $\xi$, $S_f$, $\Omega$, $n_f$ represents the solution variable, face area vector, control volume and the number of faces, respectively. The summation index $i$ represents the facets bounding the cell and the index $j$ is the vector component index (throughout).

\begin{figure}[h!]
  \begin{center}
    \includegraphics[scale=1.2]{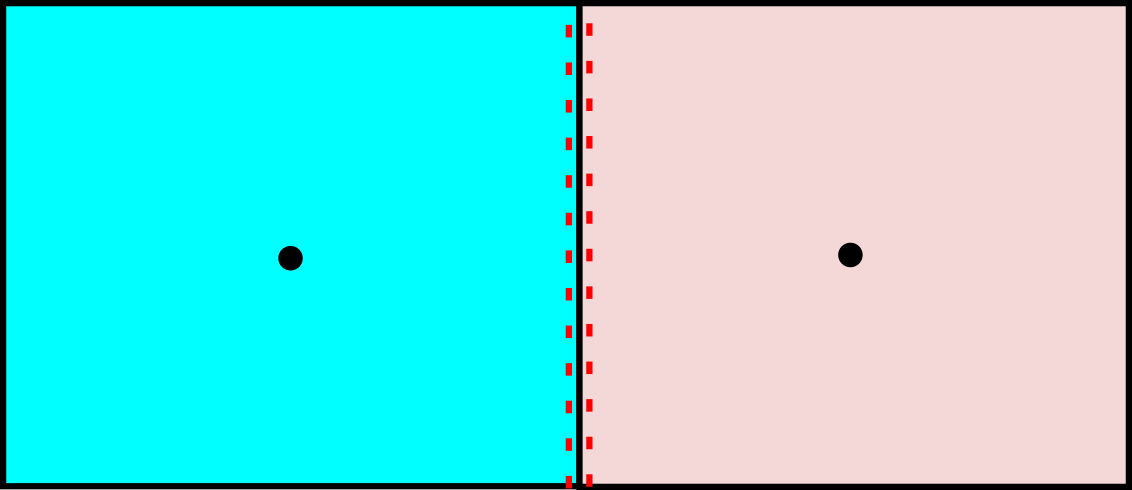}
    \label{gibContr}
    \caption{Internally, the GIB represents interface facets as two distinct boundary entities, one for each side of the surface (red dashed lines). This allows each side of the interface to interact independently with the solution matrices.}
  \end{center}
\end{figure}

Since each internal facet is connected to two cells, the replacement effected by the GIB should provide appropriate boundary contributions to both elements. To facilitate this, the GIB is split into two halves (see figure \ref{gibContr}), one for each side of the interface (like a baffle). Each half of the GIB can now provide independent boundary conditions (Derichlet, Neuman, Coupled, etc.) to the appropriate cell rather than the contributions provided by the original internal face. Revisiting the example above, the influence of the GIB boundaries can be simply expressed by:

\begin{eqnarray}   
		\int_{\Omega} \frac{\pp \xi}{\pp x_j} d\Omega
		&=&
		\sum_{i=1}^{n_f} \left({{S_f}_j} {\xi_f} \right)_i + C_{gib}\\
		C_{gib} &=&
		\sum_{k=1}^{n_{gib}} \left({{S_f}_j}  \left({\xi_{gib}}- {\xi_f}\right)\right)_k
		\label{gib_gauss} 
\end{eqnarray}

\noindent
where the index $k$ addresses the facets belonging to the same set as the GIB and the net effect of the GIB contribution $C$, is the difference between the internal face contribution and the GIB boundary condition contribution. It is simple to extent the methodology expressed in equation \ref{gib_gauss} to all other face-based operations in the finite volume framework. This will serve as a complete boundary introduction system in static mesh scenarios.

\subsection{Moving boundaries} \label{ale}

The GIB is distinct from the cut-cell method in the sense that mesh fluxes become discontinuous when a cell passes from one side of the surface to the other. The graph intersection method used to define the topology of the surface results in step changes of the interface relative to any given cell as the surface sweeps across it. Due to the large magnitude of this transitional flux and its lack of time dependence, it is difficult to treat within the framework of the ALE method (Arbitrary Lagrangian Eulerian) \cite{Hirt1974}. Consider a first order temporal discretization: the cell volumes and value of the previous iteration ($\xi^{o}$, $\Omega^{o}$) appear in the discretized form of the equation: 

\begin{eqnarray}   
        \int_{t}^{t+\Delta t}	
        \left(  	
		    \int_{\Omega} \frac{\pp \xi}{\pp t} d\Omega
		\right)    
		dt
		= 
		\frac{\xi^{n} \Omega^{n} - \xi^{o} \Omega^{o}}{\Delta t}
		\label{temporalTerm} 
\end{eqnarray}

where the super-scripts $o$ and $n$ refer to the previous and current timestep, respectively.
The history (or old time values) of elements that have transitioned from one side of the surface to the other is invalid. In fact, the only reasonable treatment is to assign an old time volume for transitional elements equal to zero. With a zero old time volume, the actual old time values held in these elements becomes irrelevant. Growing cells from zero size in a single time-step was however found to be unstable, probably due to the invalidity of the assumptions of linearity in the method. To overcome this issue, the mesh fluxes of the mesh motion are split to two parts: the transitional flux and the boundary motion flux. The transition flux is used to generate approximate old time cell volumes and old time cell values through a conservative redistribution algorithm. Once the old time quatities have been reconstructed, the boundary fluxes are employed to represent the boundary motion within the context of the ALE framework \cite{Hirt1974}.

In the ALE method, the change in volume of a cell is related to the mesh fluxes through:

\begin{eqnarray}
        \frac{\pp} {\pp t}   
        \int_{\Omega}	
		d\Omega
		&=& 
        \int_{S}
        v_s
		dS \\
		\frac{\Omega^{n} - \Omega^{o}}{\Delta t}
		&=&
		\sum_{i}^{f} {v_s}_{i} {S_f}_{i} = 
		\sum_{i}^{f} \varphi_{mesh}
		\label{scl} 
\end{eqnarray}

where ${v_s}$,  $\varphi_{mesh}$ represent the face velocity and the mesh fluxes, respectively. The mesh fluxes are calculated from the volumes swept by the faces $\Omega_{sweep}$ via:

\begin{eqnarray}
        \varphi_{mesh} =  
        \frac{\Omega_{sweep}}{\Delta t}
		\label{sweptVol} 
\end{eqnarray}

Our ALE approach splits $\varphi_{mesh}$ into the transitional part $\varphi_{tr}$, which moves elements across the interface and $\varphi_{b}$ which is a result of the surface velocity: 

\begin{eqnarray}
        \varphi_{mesh} = \varphi_{tr} + \varphi_{b}
		\label{meshFluxes} 
\end{eqnarray}

At each time-step, prior to the transport step, $\varphi_{tr}$ is used to conservatively redistribute cell properties between elements affected by the transition. All the quantities are transported while keeping the material derivative \cite{ALEbook} of the integral of the fields zero. The material derivative of the field integral is given by the following expression:

\begin{eqnarray}
        \frac {D}{Dt} \int_{\Omega} \xi d\Omega =
        \int_{\Omega} \frac{\pp \xi}{\pp t}  d\Omega +
        \int_{S} \xi \underbrace{{v_{tr}}_i n_i dS}_{\varphi_{tr}} = 0
		\label{materialDerivative} 
\end{eqnarray}

\begin{figure}[h!]
  \begin{center}
    \includegraphics[width=15cm,angle=0]{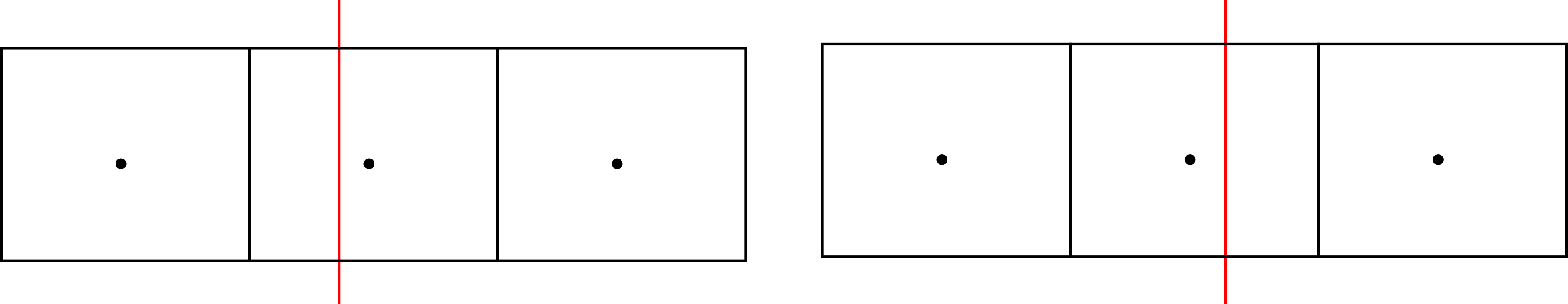}
    \caption{Simple example with a base mesh of 3 cells and a moving interface (red line) time step $n$ (left) and $n+1$ (right).}
    \label{basem}
  \end{center}
\end{figure}

After a first order accurate discretization, equation \ref{materialDerivative} takes the following form:

\begin{eqnarray}
        \frac {\xi^r \Omega^r - \xi^o \Omega^o}{\Delta t} - \sum_{i}^{n} \left(  \xi_f \varphi_{tr} \right)_i  = 0
		\label{consRedis} 
\end{eqnarray}

\noindent
where the super-script $r$ indicates the reconstructed old time value induced by pure transitional mesh fluxes, $\xi_f$ is the value of the field $\xi$ on the face and ${v_{tr}}_i$ is the mesh velocity due to the cell transition. In order to minimise the computational cost, the solution of the redistribution is performed explicitly and only on elements affected by the transition fluxes. To increase the accuracy and reduce the effects of non-linearity, sub-cycling is used for the redistribution solution. After the redistribution is complete, all old time fields and volumes have been updated to take into account the effects of the time independent transition fluxes. Then, the boundary mesh flux  $\varphi_{b}$ is used to calculate the convective flux of the state equations like a classic ALE approach:

\begin{eqnarray}
        \varphi_{rel} = \varphi_{u} - \varphi_{b}
		\label{totalFlux} 
\end{eqnarray}

\noindent
with $\varphi_{u}$ representing the absolute velocity convective flux. To better illustrate the proposed approach, a simple example of a transition cell is presented. Consider a base mesh consisting of three cells. The interface (red line) moves as depicted in figure \ref{basem}.

\begin{figure}[h!]
\centering
\subfigure[Cell transition]
{
{\includegraphics[width=7.5cm,angle=0]{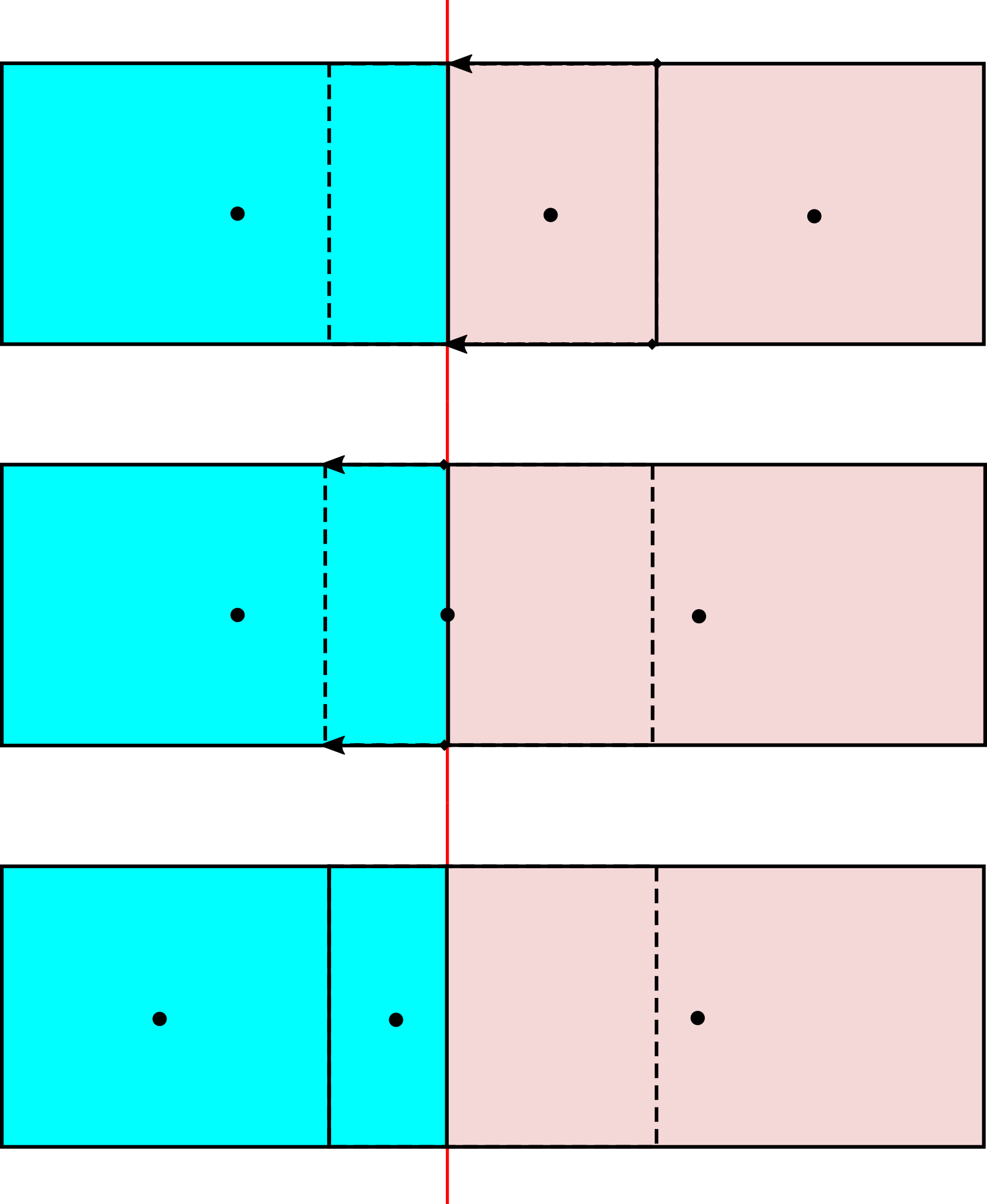}}
\label{popPict}
}
\subfigure[Conventional ALE stage]
{
{\includegraphics[width=7.5cm,angle=0]{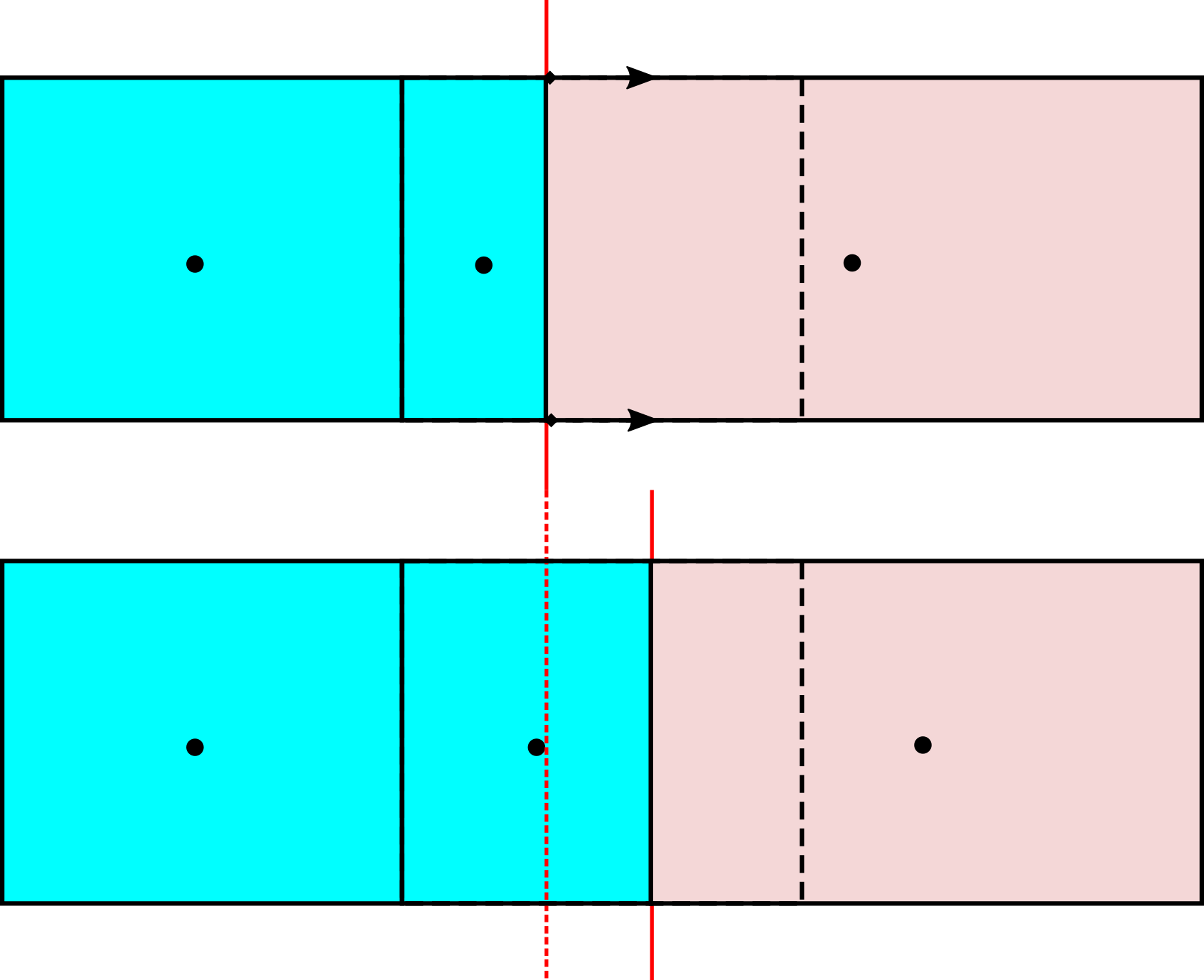}}
\label{meshPhiPict}
}
\caption {Stages of cell transition (left): first the cell shrinks to zero size and then it grows on the far side of the surface. Boundary motion after transition (right).}
\label{fluxes}
\end{figure}

When the interface passes over the cell centre of the middle cell, the cell changes "phase" (or transitions). Figure \ref{popPict} shows the two intermediate stages of the transition: 1) the transition cell is shrunk to zero size, which effectively removes all history from the element and 2) it is grown back to finite size on the opposite side of the interface. During each intermediate stage, the transition fluxes redistribute the volume and the quantities to and from the neighbouring elements. Finally, the remaining, boundary-velocity dependent, mesh fluxes (figure \ref{meshPhiPict}) are used along with the reconstructed old time values and volumes in the ALE method to fully describe the motion.

\subsubsection{Algorithm}\label{GIBAlgorithm}

The process is summarised in the following steps, which are performed prior to the solution of the governing equations:

\begin{itemize}
  \item First, the interface is moved based on the boundary velocity and the newly intersecting faces are identified. These faces are deformed such that they are coincident with the surface. The set of coincident faces is used to construct the dual sided GIB boundary.

  \item The transitional elements are identified based on the old and the new intersection graph.
  
  \item The mesh fluxes,  $\varphi_{mesh}$, are calculated, using the old and the new point coordinates of the mesh.
  
  \item The transition $\varphi_{tr}$ and the boundary $\varphi_{b}$ fluxes are calculated based on the boundary velocity of the interface.
  
  \item The old GIB boundary values of all the fields are mapped to the new GIB boundaries using consistent surface to surface interpolation.
  
  \item Transported fields are conservatively redistributed to take into account transitional fluxes (pure mesh motion). 
  
  \item The boundary motion is incorporated via the conventional ALE approach.
  
\end{itemize}

\vspace{20 mm}

\section{Validation}\label{}

\begin{figure}[h!]
\centering
\subfigure[Computational mesh overview]
{
{\includegraphics[height=7cm,angle=0]{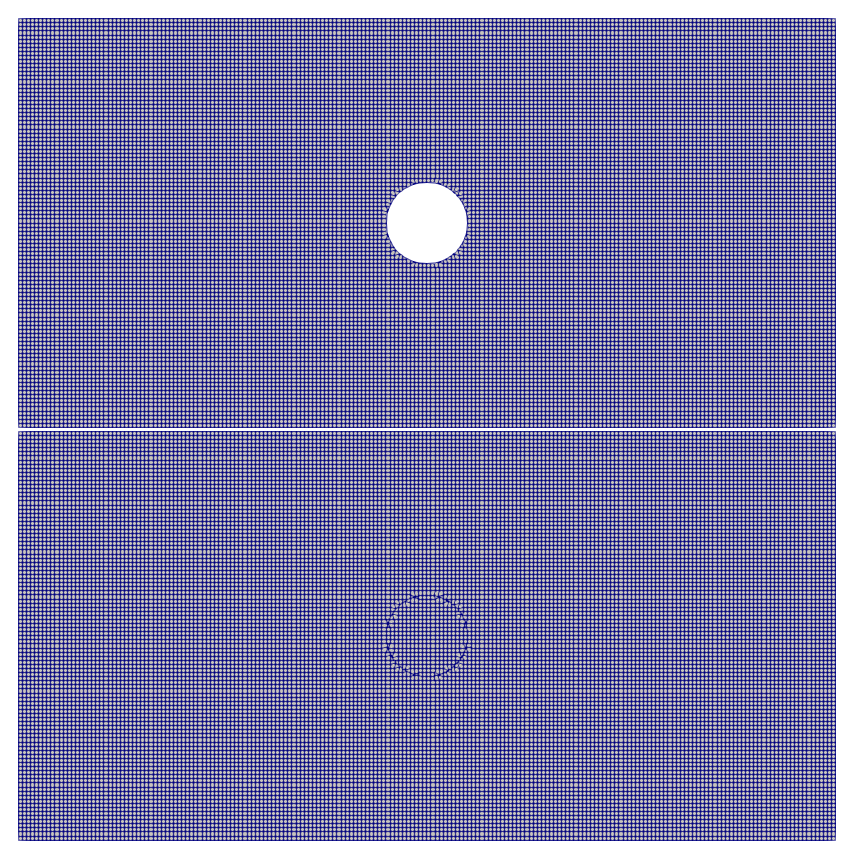}}
\label{val:cylinder:mesh1}
}
\subfigure[Computational mesh in close proximity to the cylinder]
{
{\includegraphics[height=7cm,angle=0]{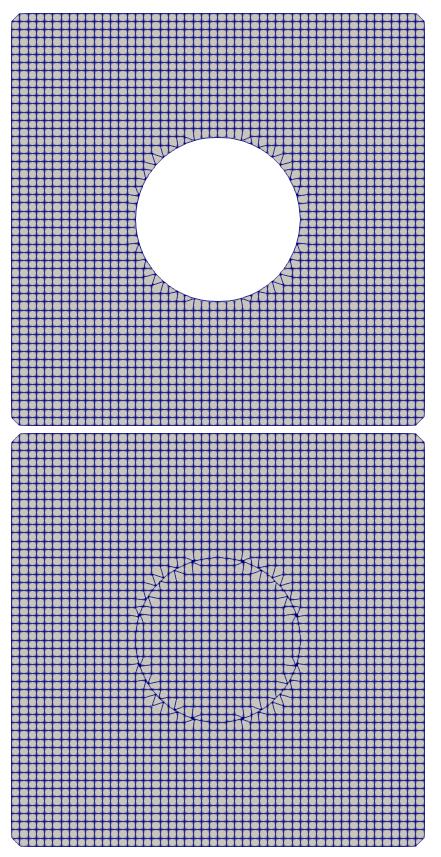}}
\label{val:cylinder:mesh2}
}
\caption {Computational mesh of the cylinder with GIB (bottom) and peripheral boundaries (top). The flow enters the computational domain from the left (inlet) and it exits from the right (outlet). All the other boundaries are no-slip walls.}
\label{val:cylinder}
\end{figure}

In this section, the GIB approach is validated against a body fitted mesh for a simple two-dimensional cylinder (figure \ref{val:cylinder}). In the first stage, the cylinder is static to ensure the proposed method produces the same results as the body fitted approach (section \ref{static}). In the second stage, horizontal and vertical oscillating motion is added to the cylinder (section \ref{moving}). Finally, the GIB are validated against experimental results for a butterfly valve (section \ref{valve}).

\subsection{Static Cylinder}\label{static}

\begin{figure}[h!]
\centering
\subfigure[Velocity field]
{
{\includegraphics[width=6cm,angle=0]{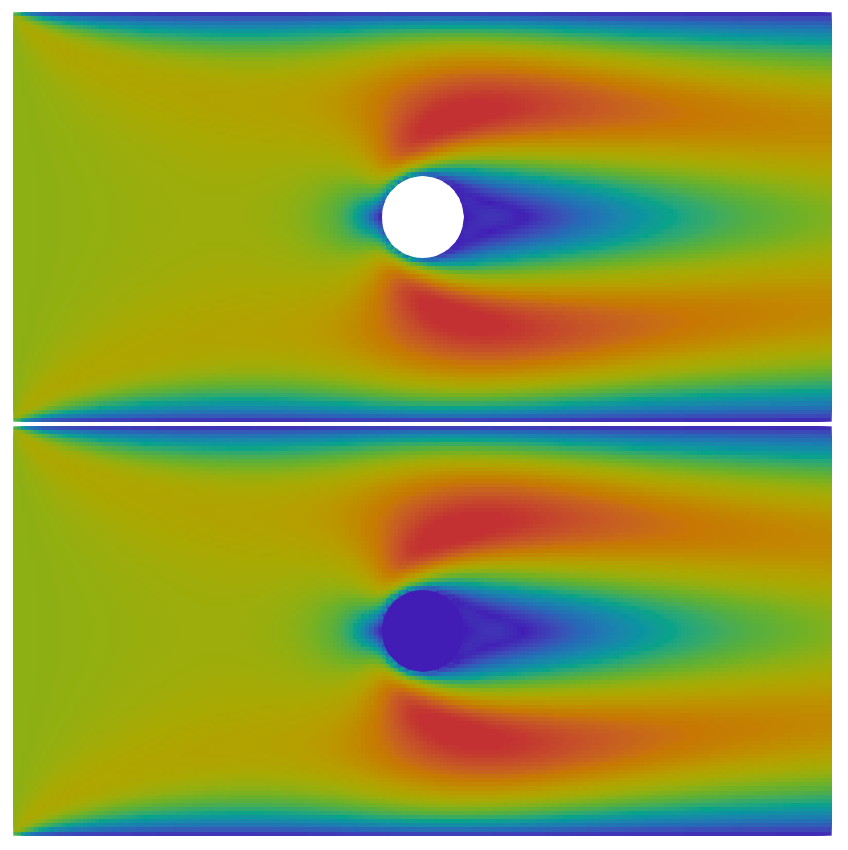}}
\label{val:static:U}
}
\subfigure[Pressure field]
{
{\includegraphics[width=6cm,angle=0]{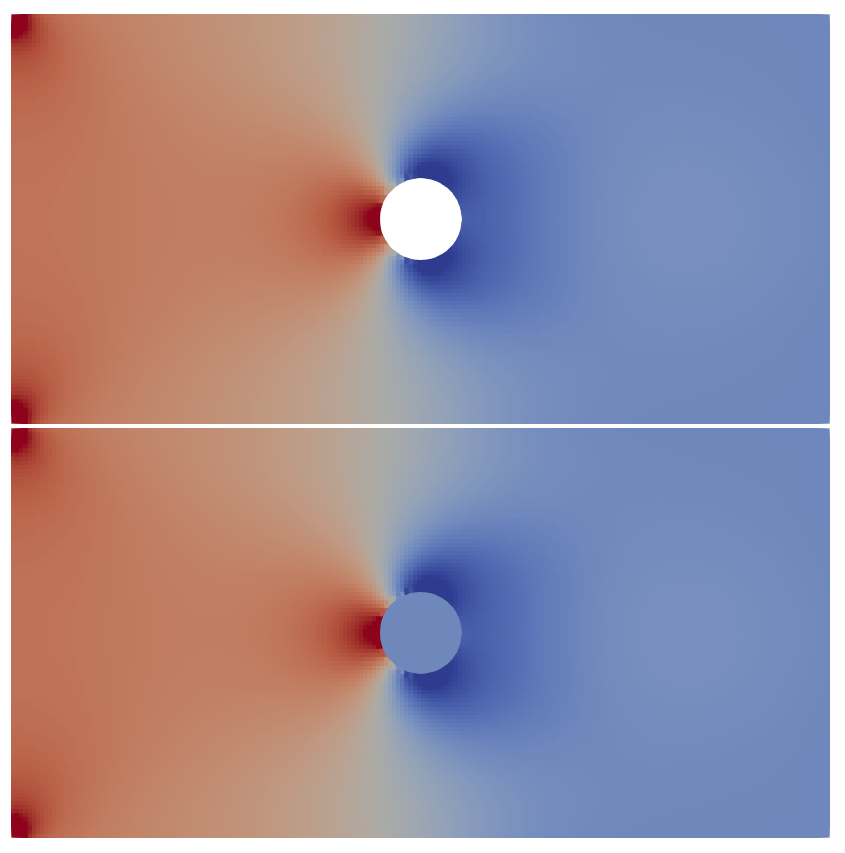}}
\label{val:static:p}
}
\caption {Static cylinder result (pressure and velocity): Body-fitted (top) GIB (bottom).}
\label{val:static}
\end{figure}

\begin{figure}[h!]
  \begin{center}
    \includegraphics[scale=0.6]{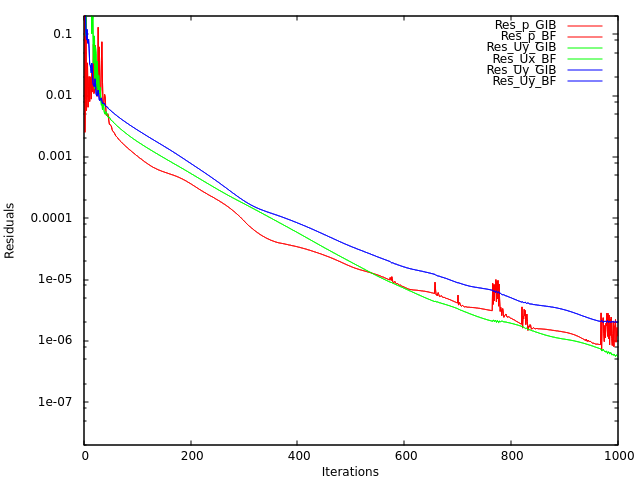}
    \caption{Velocity and pressure residuals of the body-fitted and the GIB case.}
    \label{val:static:res}
  \end{center}
\end{figure}

The governing equations for the static comparison are the the steady incompressible Navier-Stokes equations. The $SIMPLE$ algorithm \cite{patankar1980numerical,ferziger2001} is used for the solution of the state equations. Figure \ref{val:static} compares the velocity and pressure fields on the two cases. The results are indistinguishable, other than the solution field that is visible on the inside of the cylinder in the GIB case. Similarly, the residuals for pressure and velocity in the two cases are identical (figure \ref{val:static:res}). 

\begin{figure}[h!]
\centering
\subfigure[Position of probes in the mesh]
{
{\includegraphics[width=5cm,angle=0]{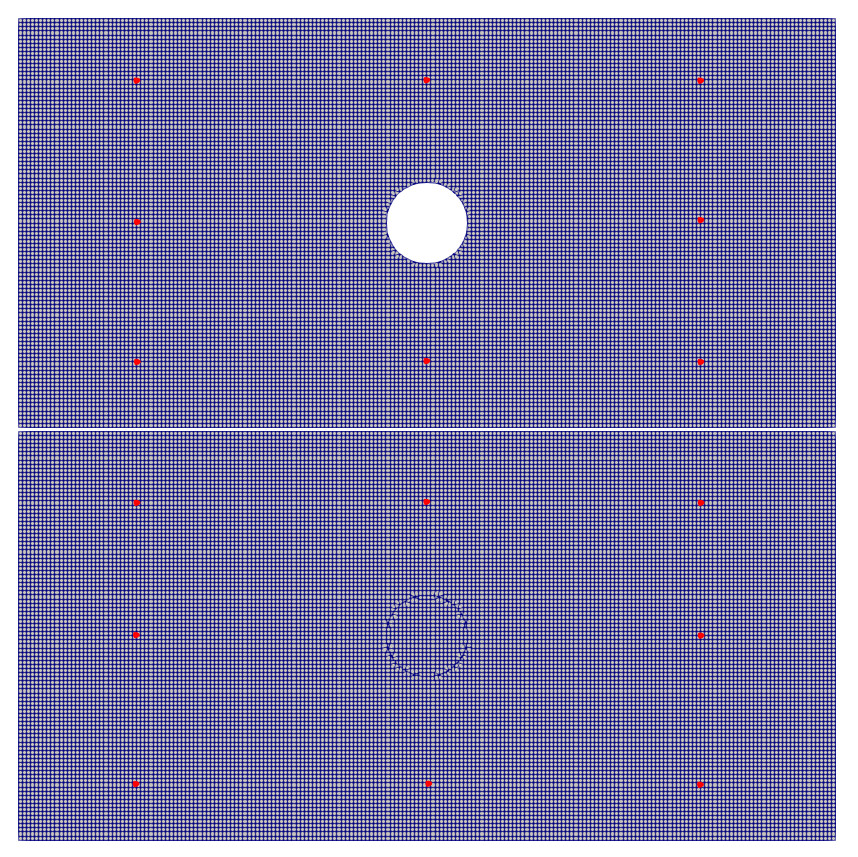}}
\label{val:static:probes:mesh}
}
\subfigure[Difference in velocity magnitude (equation \ref{diffU}) per iteration]
{
{\includegraphics[width=6.5cm,angle=0]{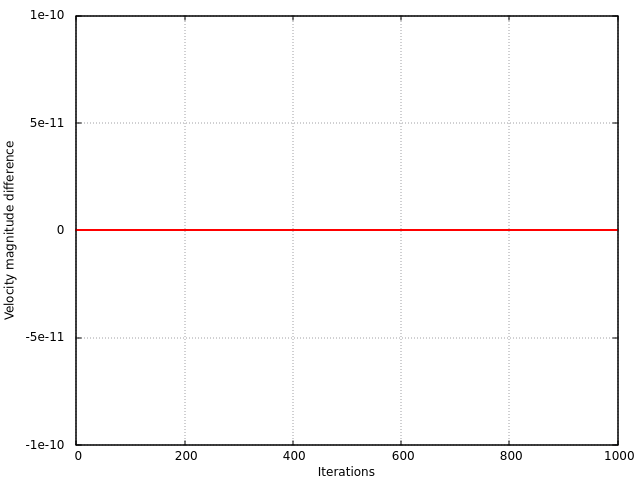}}
\label{val:static:probes:U}
}
\caption {Comparison of velocities between body-fitted and GIB meshes at 8 probe locations.}
\label{val:static:probes}
\end{figure}

In order to ensure that the two cases produce identical results, the difference in velocity is calculated (equation \ref{diffU}) at discrete points in the fluid domain and depicted in figure \ref{val:static:probes:mesh}. As expected, the results in the two cases are identical (figure \ref{val:static:probes}). The combination of results indicates that, for this case at least, the representation of the boundary provided by the GIB is identical in every way to that provided by the peripheral boundary. Again, this is totally expected, as, outside of the sub-domain inside the GIB cylinder, the geometry and matrices for both cases should be exactly the same.

\begin{eqnarray}
        \Delta v_{probes} = \sum_{i}^{nProbes=8} 
            \left| {v_{GIB}}_i - {v_{BF}}_i  \right|
 		\label{diffU} 
\end{eqnarray}

\subsection{Moving Cylinder}\label{moving}
In order to test the GIB performance as a moving boundary, oscillatory vertical and horizontal motion is imposed on the cylinder boundaries. The motion of the cylinder is  prescribed by the following expression:

\begin{eqnarray}
        x \left( t \right) = A sin 
          \left( 
              \omega \left( t-t_0 \right)
          \right)
		\label{motion} 
\end{eqnarray}

where $A$, $\omega$, $t$, $t_0$ are the amplitude, angular velocity, time and initialization time, respectively. The velocity of the boundary can be derived through differentiation from eq. \ref{motion} as $v(t)=dx/dt$. However, to avoid inconsistent fluxes at the boundaries from the exact and discretized forms of the velocity, the normal component of velocity must be set equal to the boundary mesh flux, $\varphi_b$. The unsteady Navier-Stokes equations are solved using the $PISO$ algorithm of Issa \cite{IssaPISO}. 

\subsubsection{Horizontial motion}\label{hormoving}

We first consider horizontal motion of the cylinder in isolation. The following parameters are chosen (equation \ref{motion}): $A=(1, 0 , 0) m$ , $\omega = 0.25132 rad/s$ (cycle period $T=25s$). The total simulation time is 4 oscillation periods (100 sec).

\begin{figure}[h!]
  \begin{center}
    \includegraphics[scale=0.8]{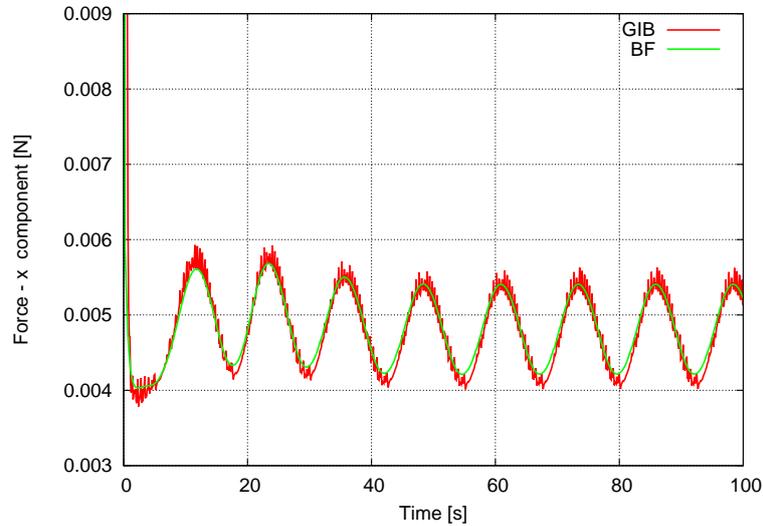}
    \caption{This figure represents the x component of the forces per time. The oscillations appeared in the GIB are caused from mesh dependency due to the coarseness of the mesh.}
    \label{val:vmoving:forces}
  \end{center}
\end{figure}

In figures \ref{val:vmoving:mesh} and \ref{val:vmoving:U} the mesh and the velocities in four different timesteps (t=10, 30, 50, 70) are presented for the body-fitted and the GIB cases. The comparison of velocity fields (figure \ref{val:vmoving:U}) and forces (figure \ref{val:vmoving:forces}) show similar results, with minor deviations.

\begin{figure}[h!]
\centering
\subfigure[t=10s]
{
{\includegraphics[width=3.5cm,angle=0]{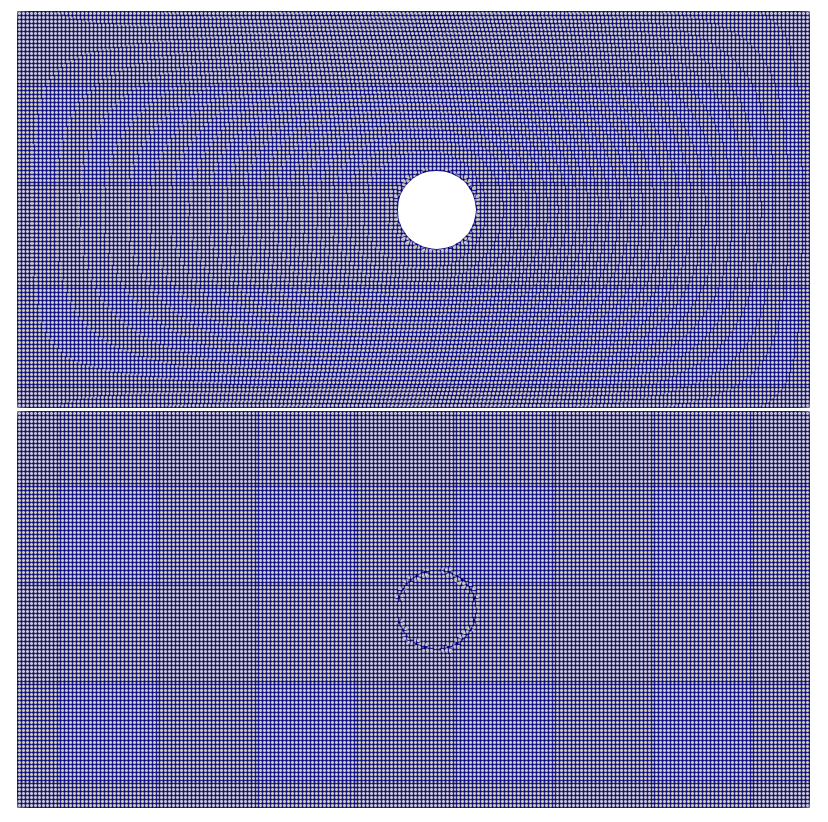}}
\label{val:hmoving:mesh10}
}
\subfigure[t=30s]
{
{\includegraphics[width=3.5cm,angle=0]{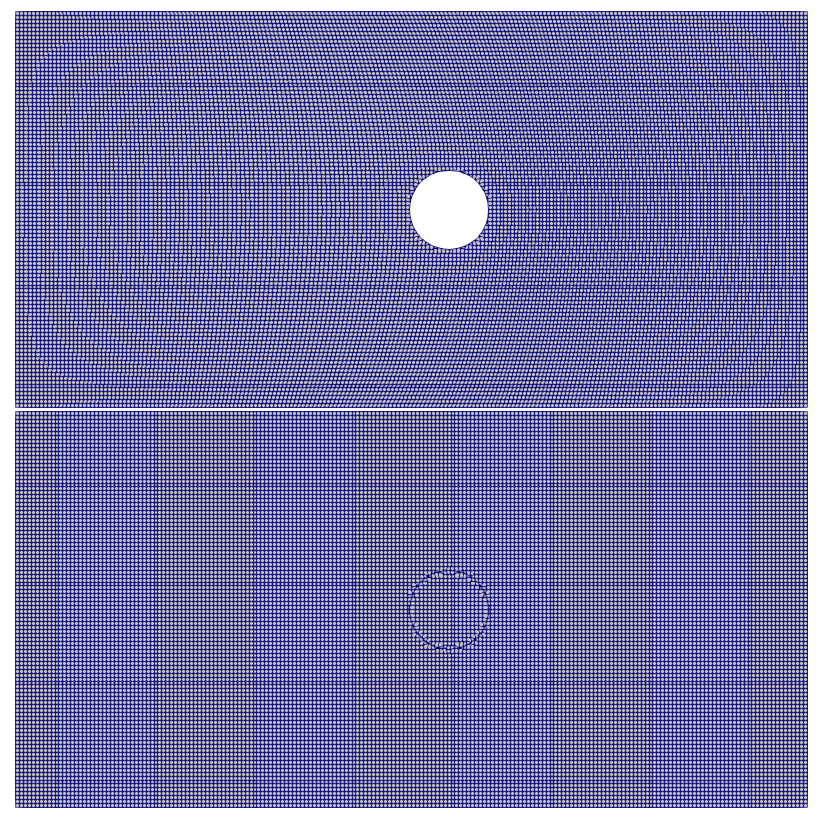}}
\label{val:hmoving:mesh30}
}
\subfigure[t=50s]
{
{\includegraphics[width=3.5cm,angle=0]{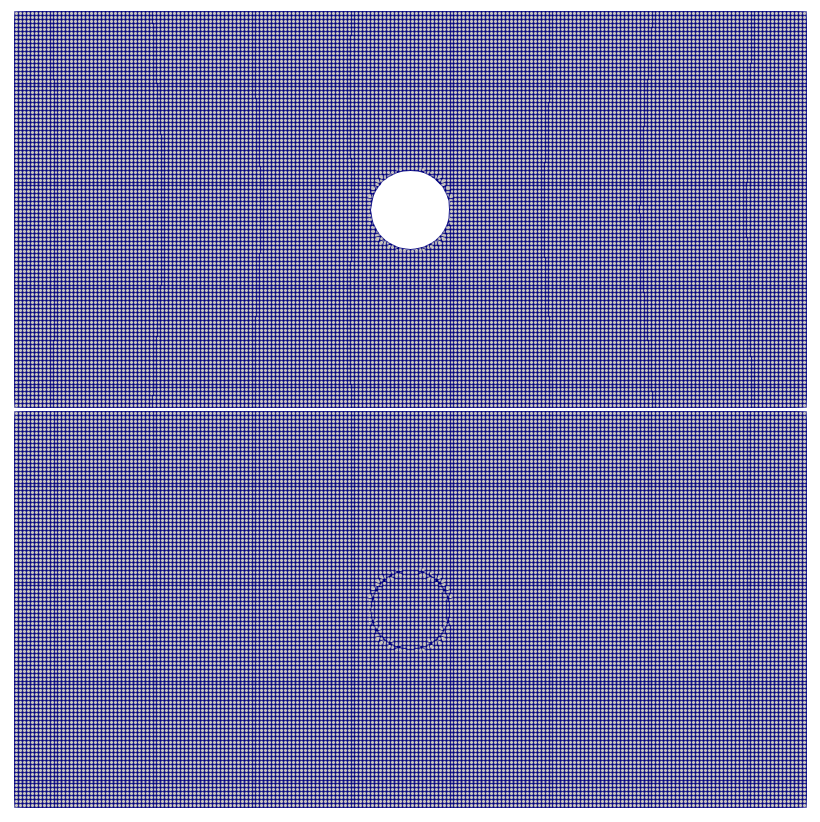}}
\label{val:hmoving:mesh50}
}
\subfigure[t=70s]
{
{\includegraphics[width=3.5cm,angle=0]{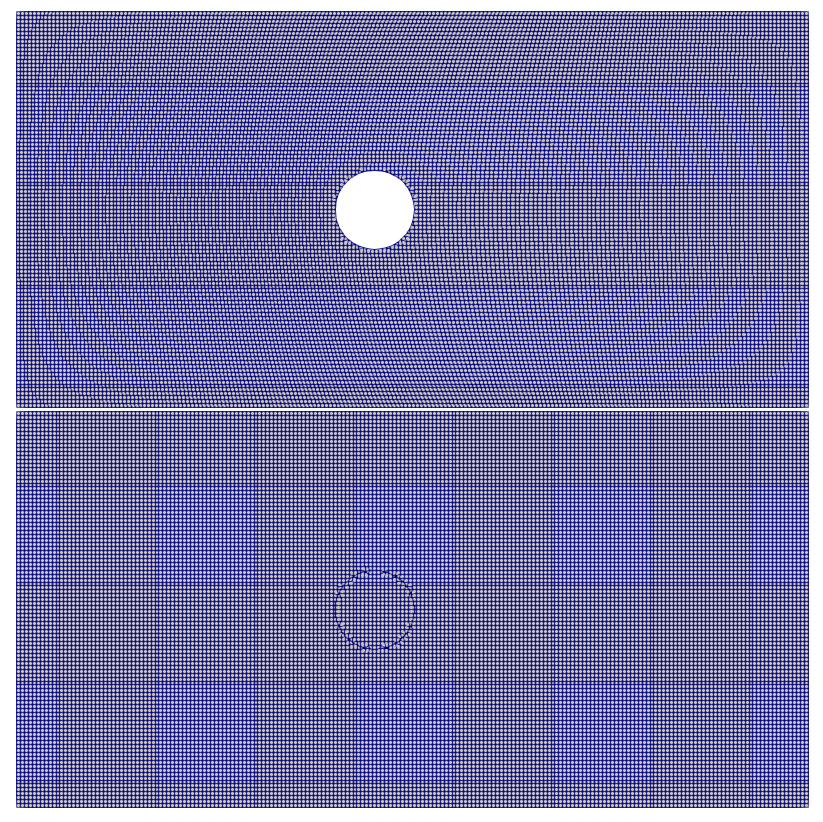}}
\label{val:hmoving:mesh70}
}
\caption {Computational mesh of GIB (bottom) and body-fitted (top) cases for four different positions}
\label{val:hoving:mesh}
\end{figure}

\begin{figure}[h!]
\centering
\subfigure[t=10s]
{
{\includegraphics[width=3.5cm,angle=0]{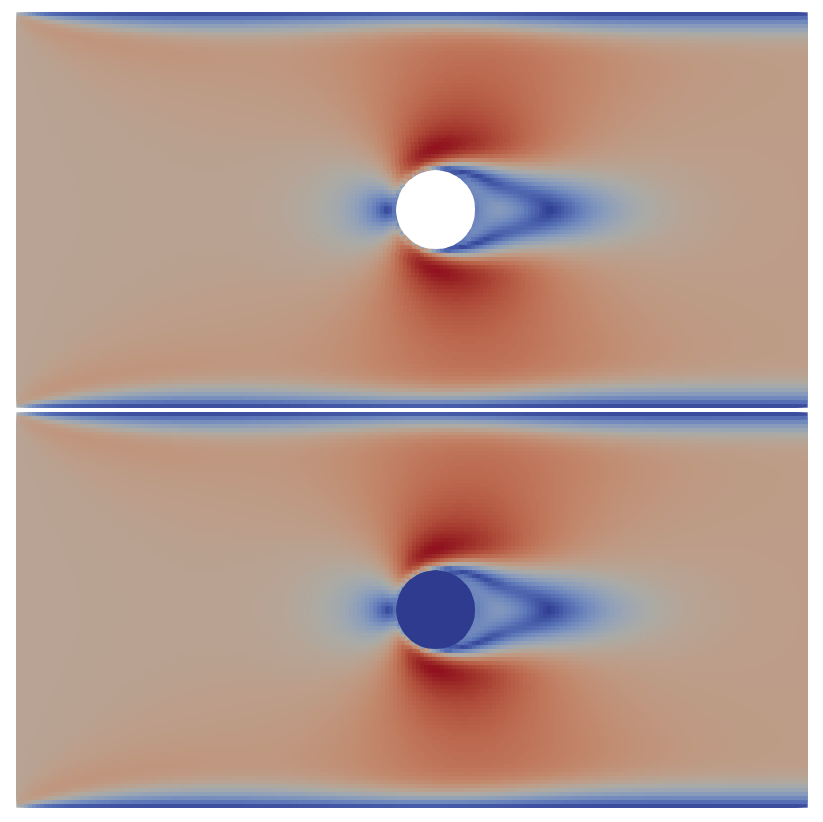}}
\label{val:hmoving:U10}
}
\subfigure[t=30s]
{
{\includegraphics[width=3.5cm,angle=0]{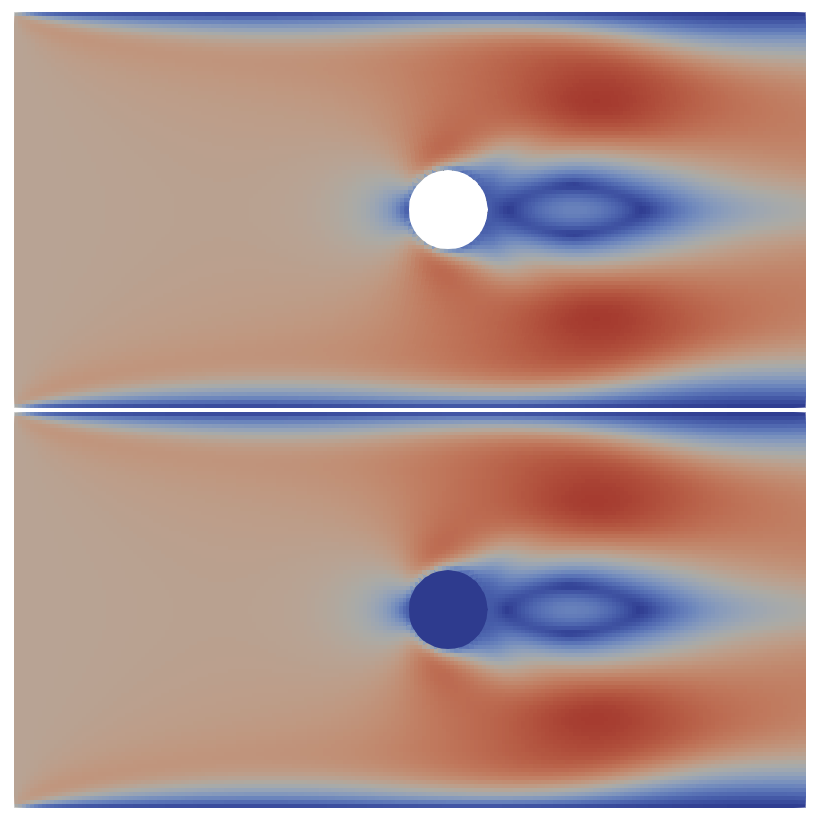}}
\label{val:hmoving:U30}
}
\subfigure[t=50s]
{
{\includegraphics[width=3.5cm,angle=0]{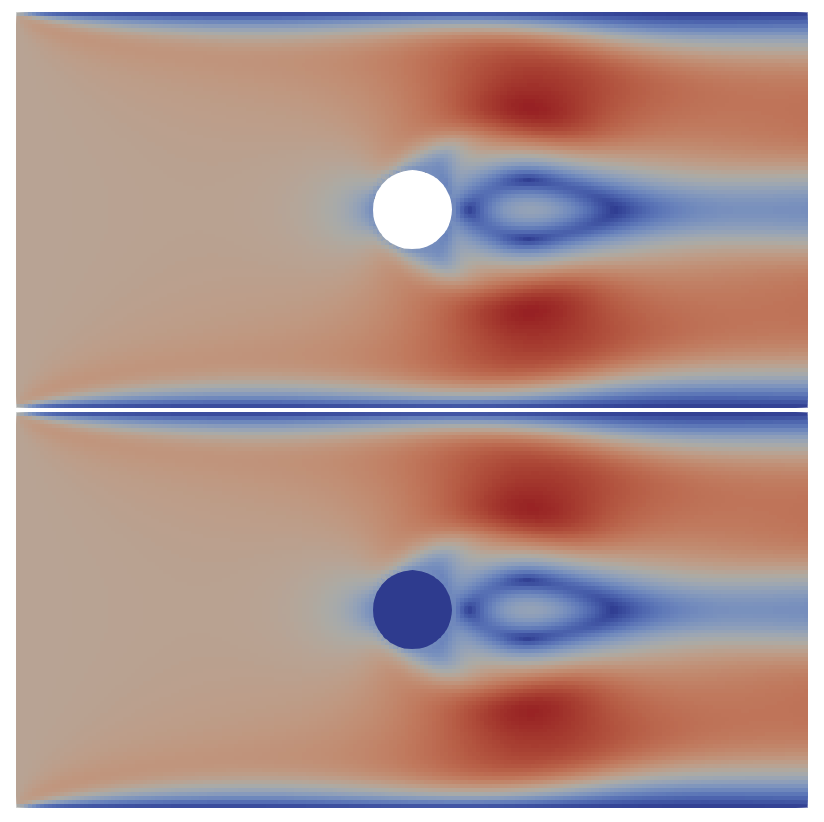}}
\label{val:hmoving:U50}
}
\subfigure[t=70s]
{
{\includegraphics[width=3.5cm,angle=0]{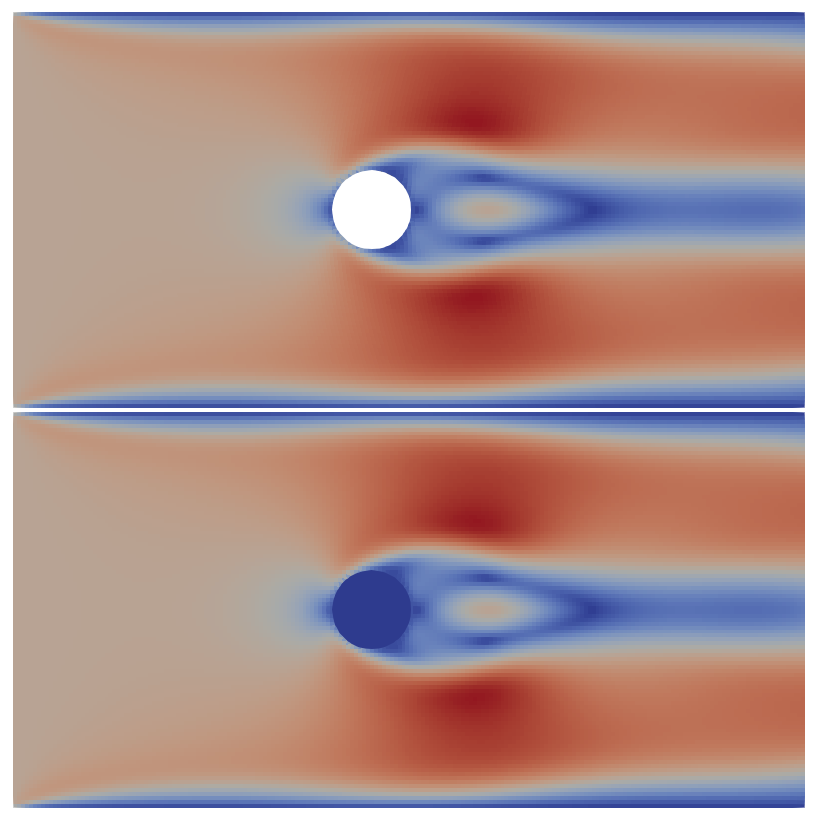}}
\label{val:hmoving:U70}
}
\caption {Velocity magnitude for the GIB (bottom) and body-fitted (top) cases for four different timesteps.}
\label{val:hmoving:U}
\end{figure}

\begin{figure}[h!]
  \begin{center}
    \includegraphics[scale=0.75]{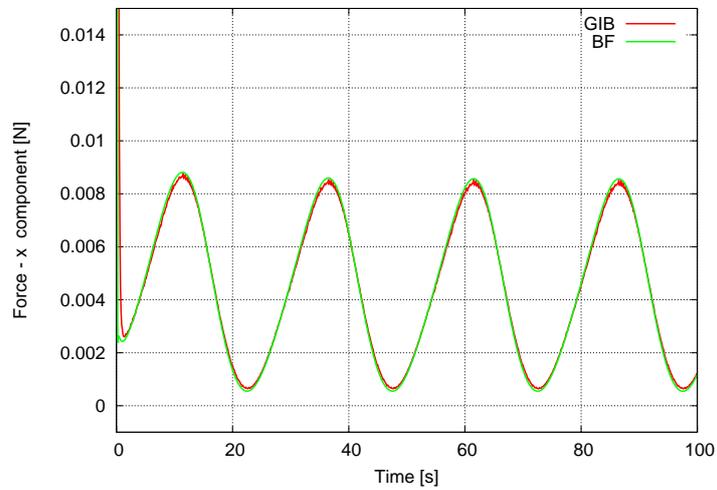}
    \caption{X component of forces vs. time.}
    \label{val:static::res}
  \end{center}
\end{figure}

\newpage

\subsubsection{Vertical motion}\label{vermoving}
\begin{figure}[h!]
\centering
\subfigure[t=10 s]
{
{\includegraphics[width=3.5cm,angle=0]{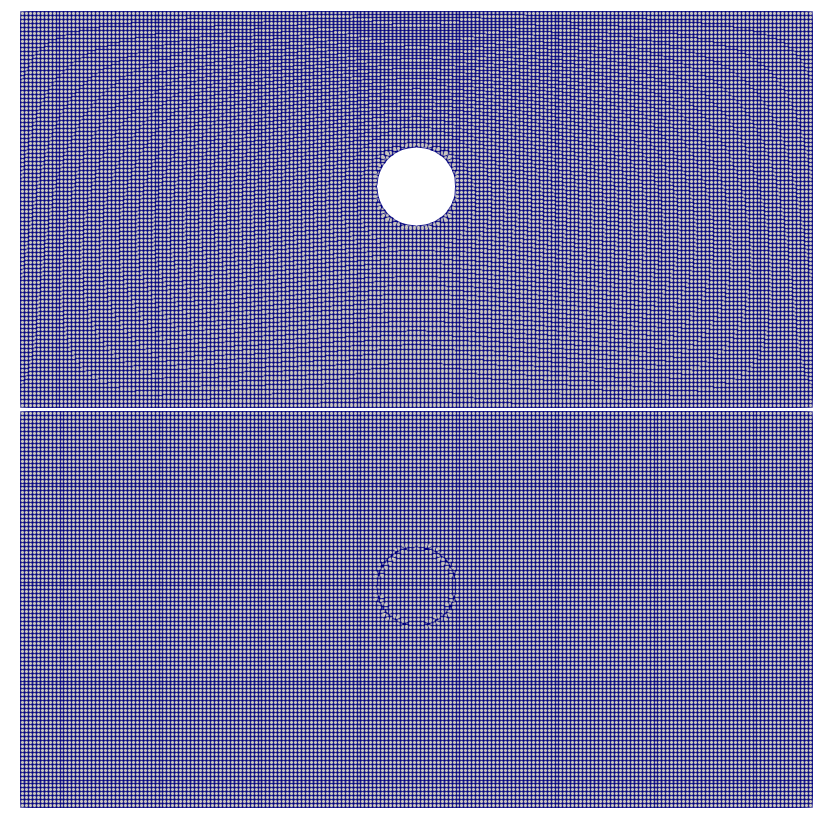}}
\label{val:vmoving:mesh10}
}
\subfigure[t=30 s]
{
{\includegraphics[width=3.5cm,angle=0]{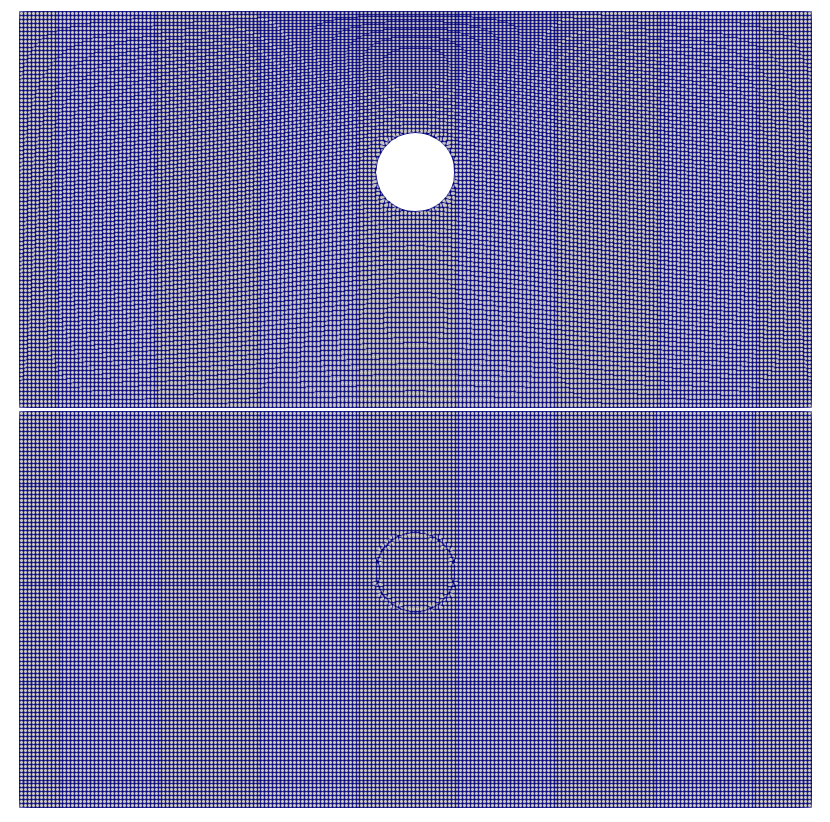}}
\label{val:vmoving:mesh30}
}
\subfigure[t=50 s]
{
{\includegraphics[width=3.5cm,angle=0]{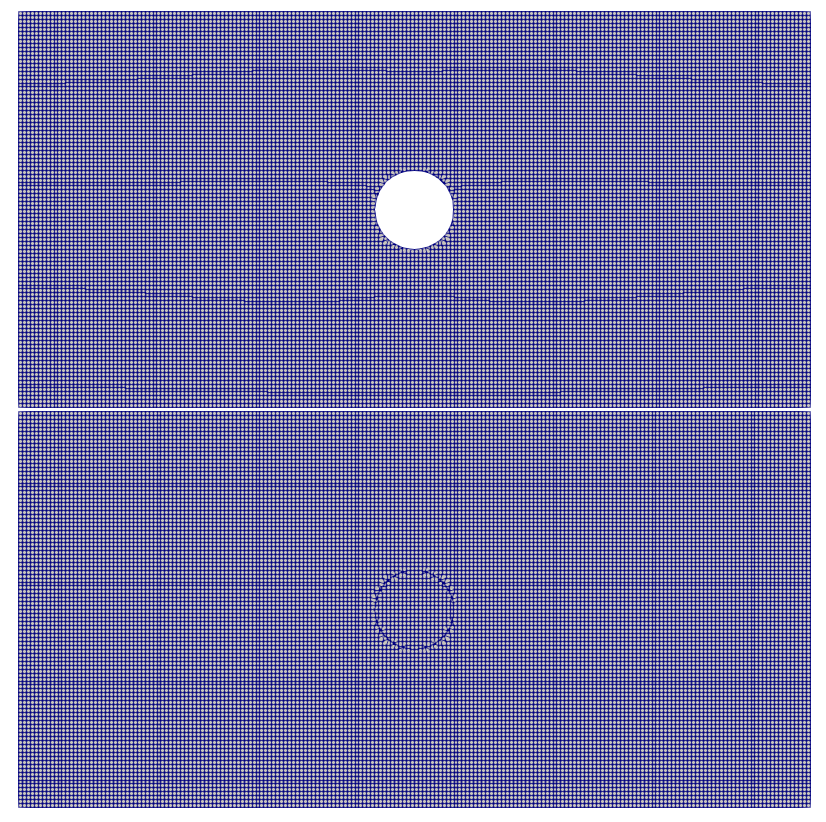}}
\label{val:vmoving:mesh50}
}
\subfigure[t=70 s]
{
{\includegraphics[width=3.5cm,angle=0]{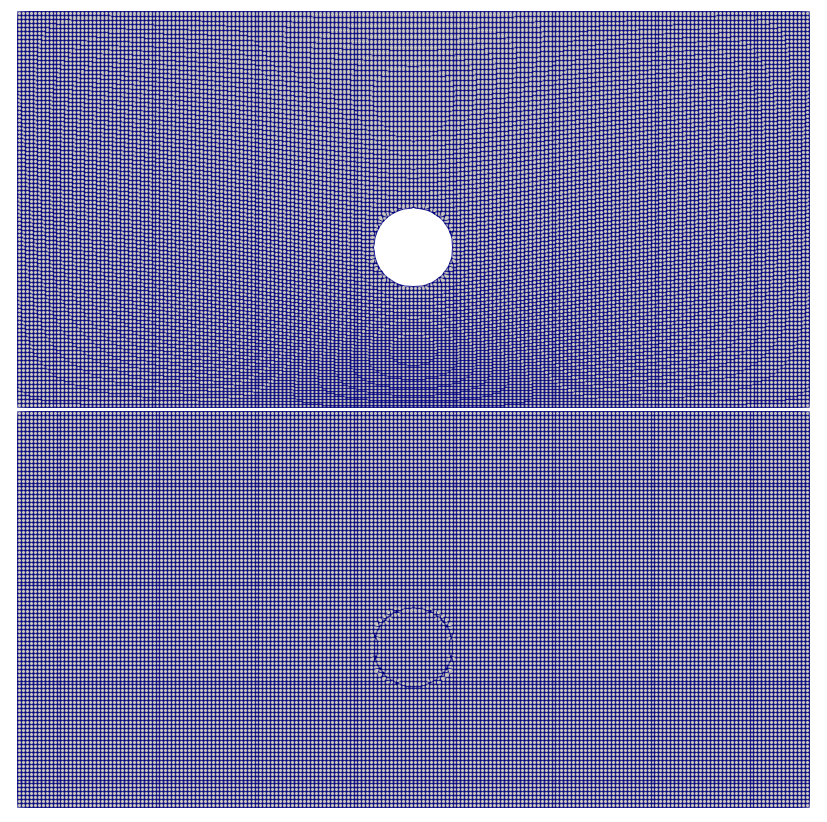}}
\label{val:vmoving:mesh70}
}
\caption {Computational mesh of GIB (bottom) and body-fitted (top) case for four different positions}
\label{val:vmoving:mesh}
\end{figure}

\begin{figure}[h!]
\centering
\subfigure[t=10 s]
{
{\includegraphics[width=3.5cm,angle=0]{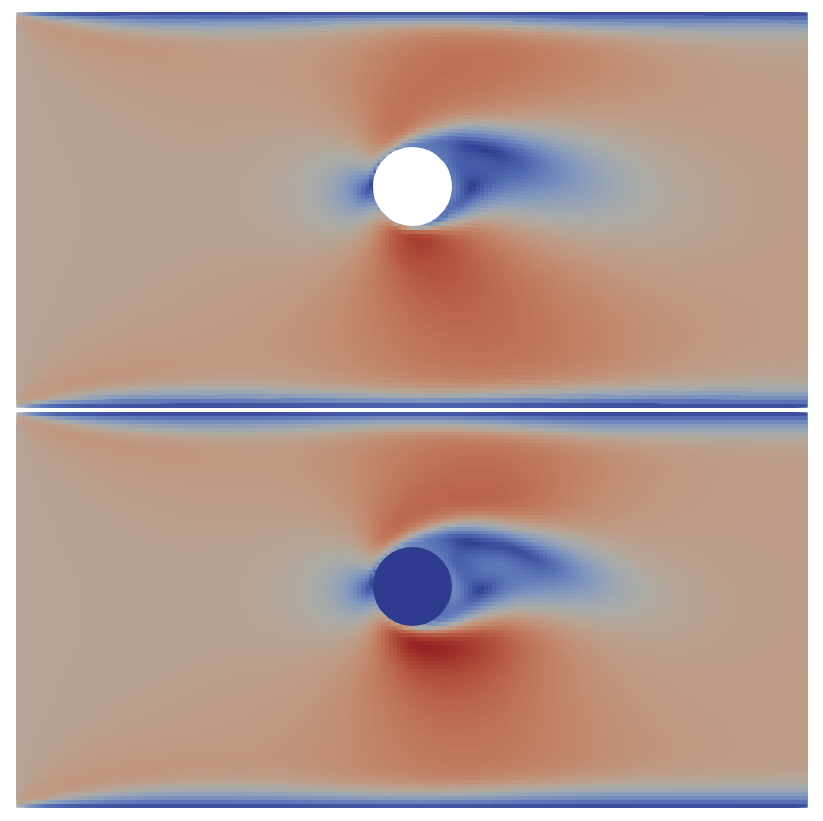}}
\label{val:vmoving:U10}
}
\subfigure[t=30 s]
{
{\includegraphics[width=3.5cm,angle=0]{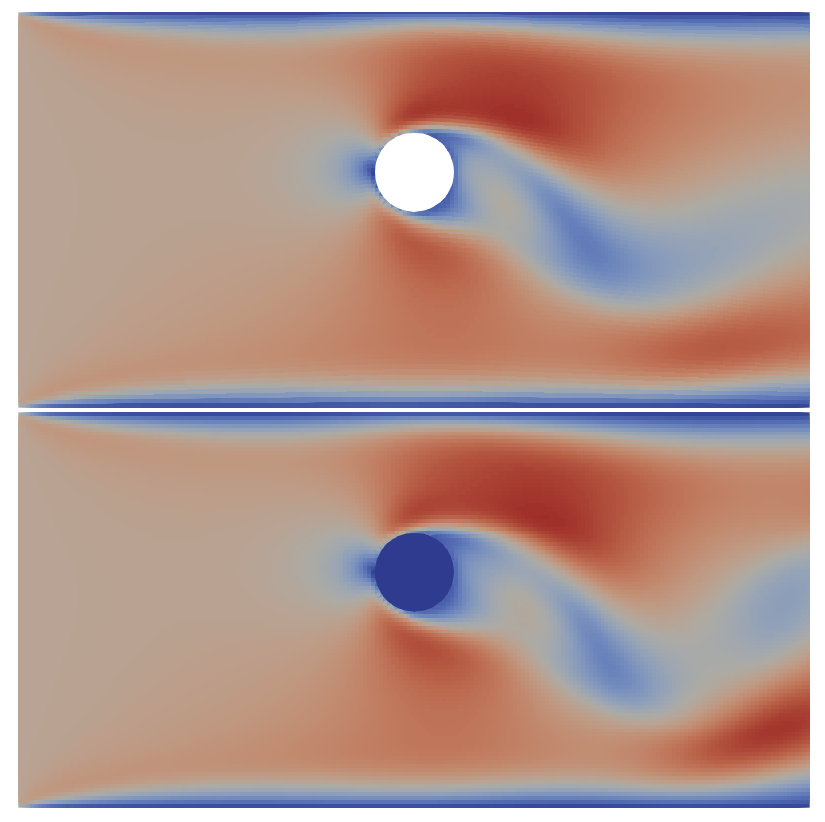}}
\label{val:vmoving:U30}
}
\subfigure[t=50 s]
{
{\includegraphics[width=3.5cm,angle=0]{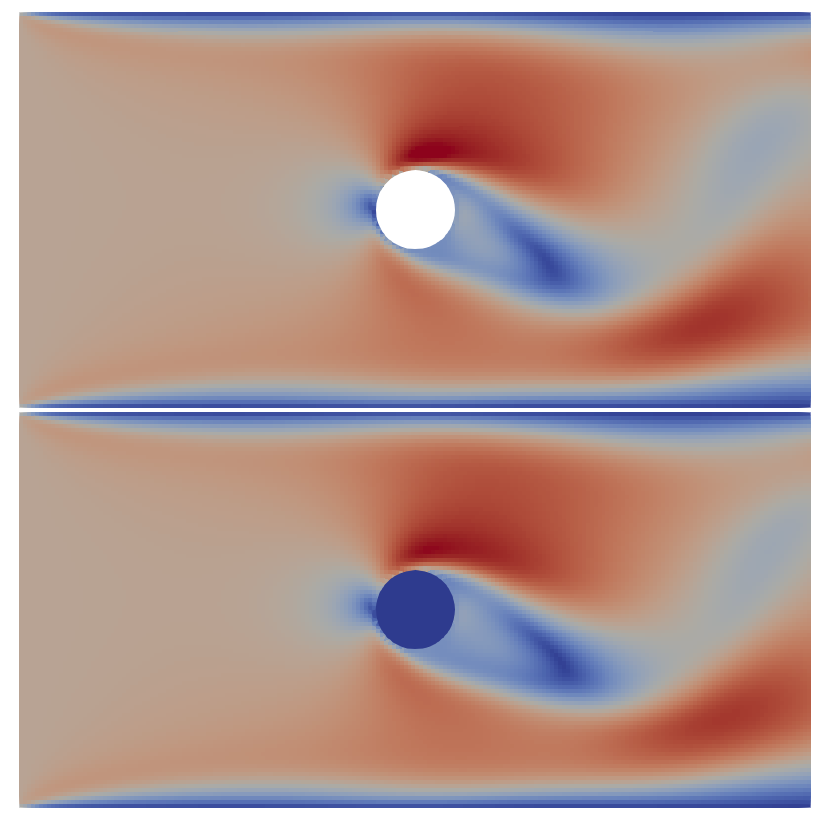}}
\label{val:vmoving:U50}
}
\subfigure[t=70 s]
{
{\includegraphics[width=3.5cm,angle=0]{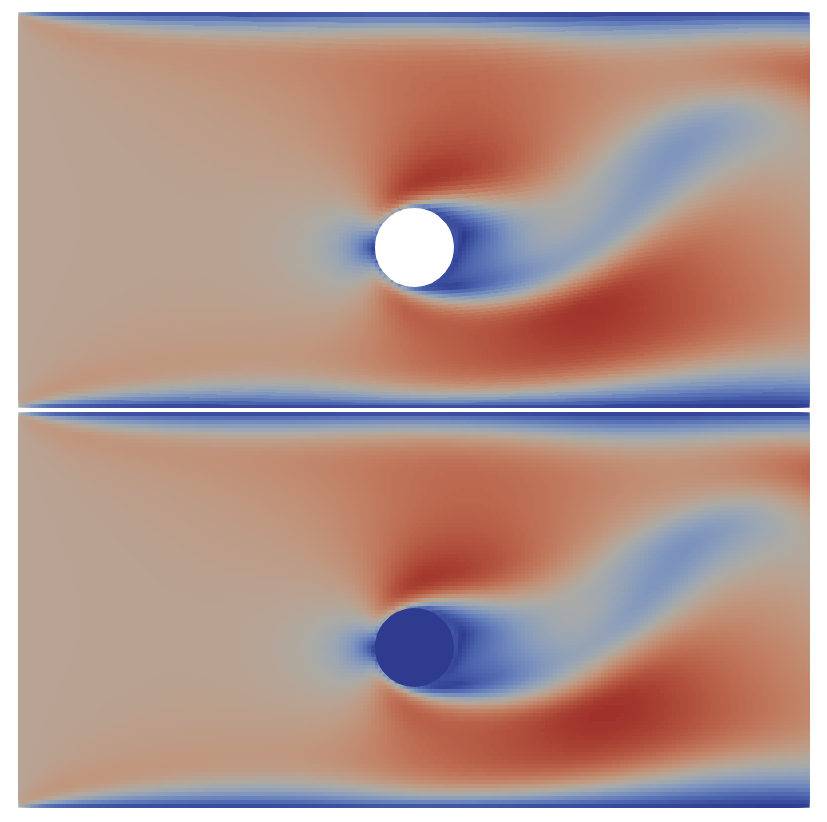}}
\label{val:vmoving:U70}
}
\caption {Velocity magnitude for the GIB (bottom) and body-fitted (top) case for four different timesteps.}
\label{val:vmoving:U}
\end{figure}

The second dynamic mesh comparison is about the vertical motion of the cylinder, so that the motion is normal to the oncoming flow. Figures \ref{val:vmoving:mesh} and \ref{val:vmoving:U} again show the mesh and the velocities for the four different timesteps (t=10,30,50,70) for the body fitted and the GIB cases. Although the velocity fields are comparable, we can see a distinct high frequency oscillatory signal in the GIB force calculations.

\begin{figure}[h!]
  \begin{center}
    \includegraphics[scale=0.6]{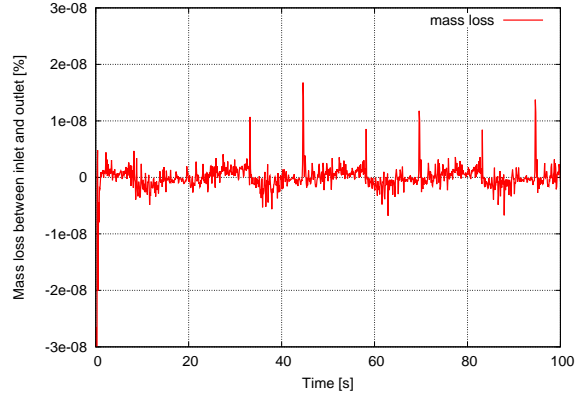}
    \caption{Mass loss percentage between inlet and outlet per iteration.}
    \label{val:vmoving:massConservation}
  \end{center}
\end{figure}

The main difference between the horizontal motion case and the case with vertical motion is the fact that the mesh resolution with respect to the high gradient shear layer on the top and bottom of the cylinder changes considerable in the vertical case, while it remaining effectively constant for the horizontal case. Unlike traditional mesh motion strategies, all the deformation in the GIB approach is clusters right next to the boundary. During cell transition, the mesh on the donor side of the interface is instantaneously coarsened  in the direction normal to the surface by a factor of 1:3. Conversely, the acceptor side sees an inverse increase in resolution. These changes in resolution will produces oscillations in the force balance if the base mesh is not fine enough to accurately resolve all non-linearities in the field. Moreover, while the conservative redistribution of the fields using the mesh fluxes is a reasonable approximation of the old time fields, the updated fields are not guaranteed to satisfy the governing equations. We would thus expect the magnitude of the oscillations to decrease with increased resolution and finally disappear as the mesh is refined. Figure \ref{val:vmoving:massConservation} shows that the mass loss between inlet and outlet per iteration is insignificant, which is expected because the presented method is mass conservative. 

\subsubsection{Mesh dependency}\label{vermoving_meshDependency}

\begin{figure}[h!]
\centering
{
{\includegraphics[width=3.5cm,angle=0]{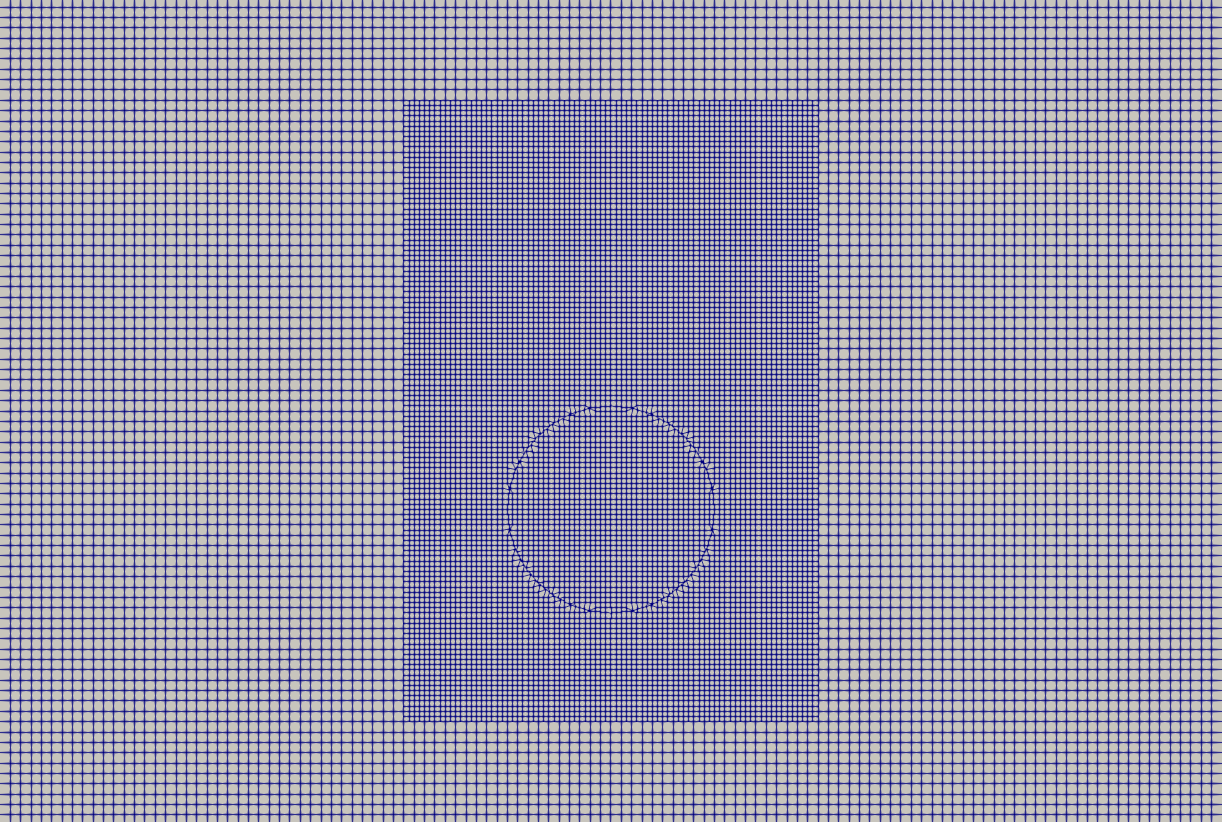}}
}
{
{\includegraphics[width=3.5cm,angle=0]{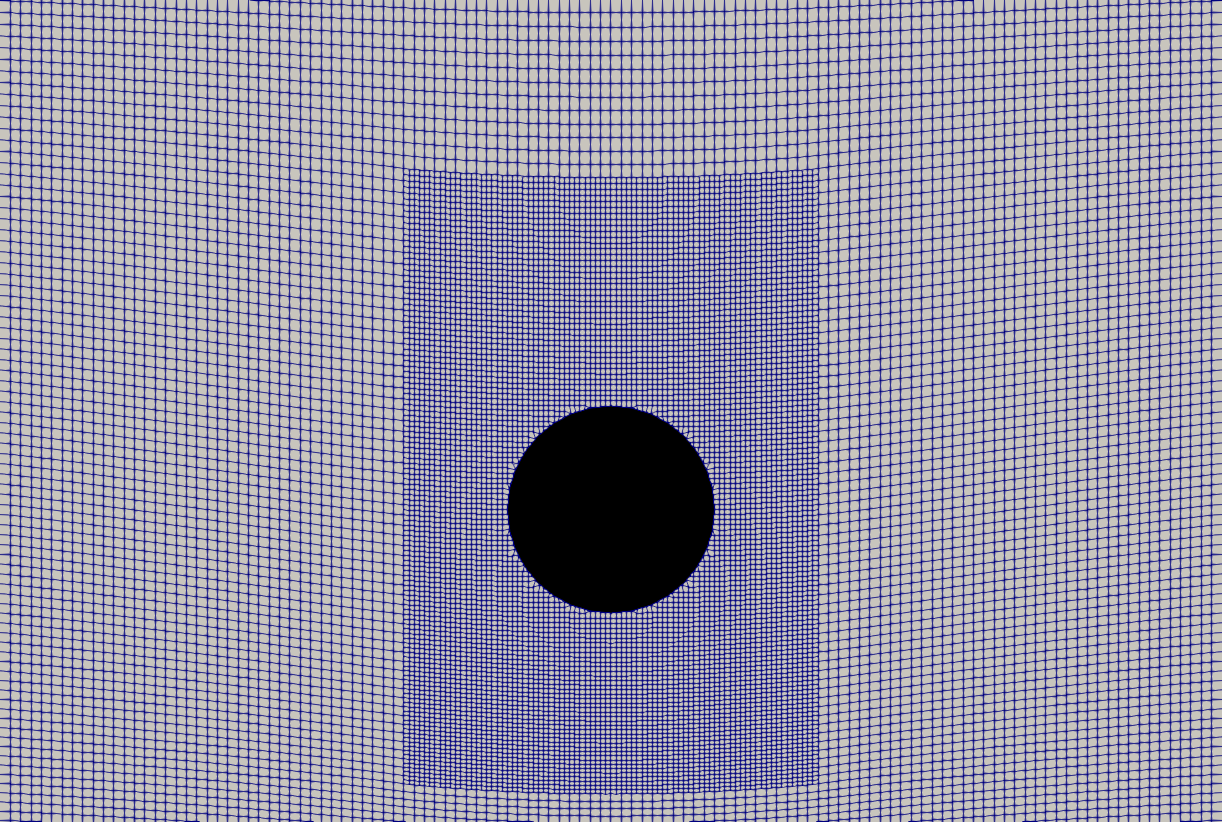}}
}
{
{\includegraphics[width=3.5cm,angle=0]{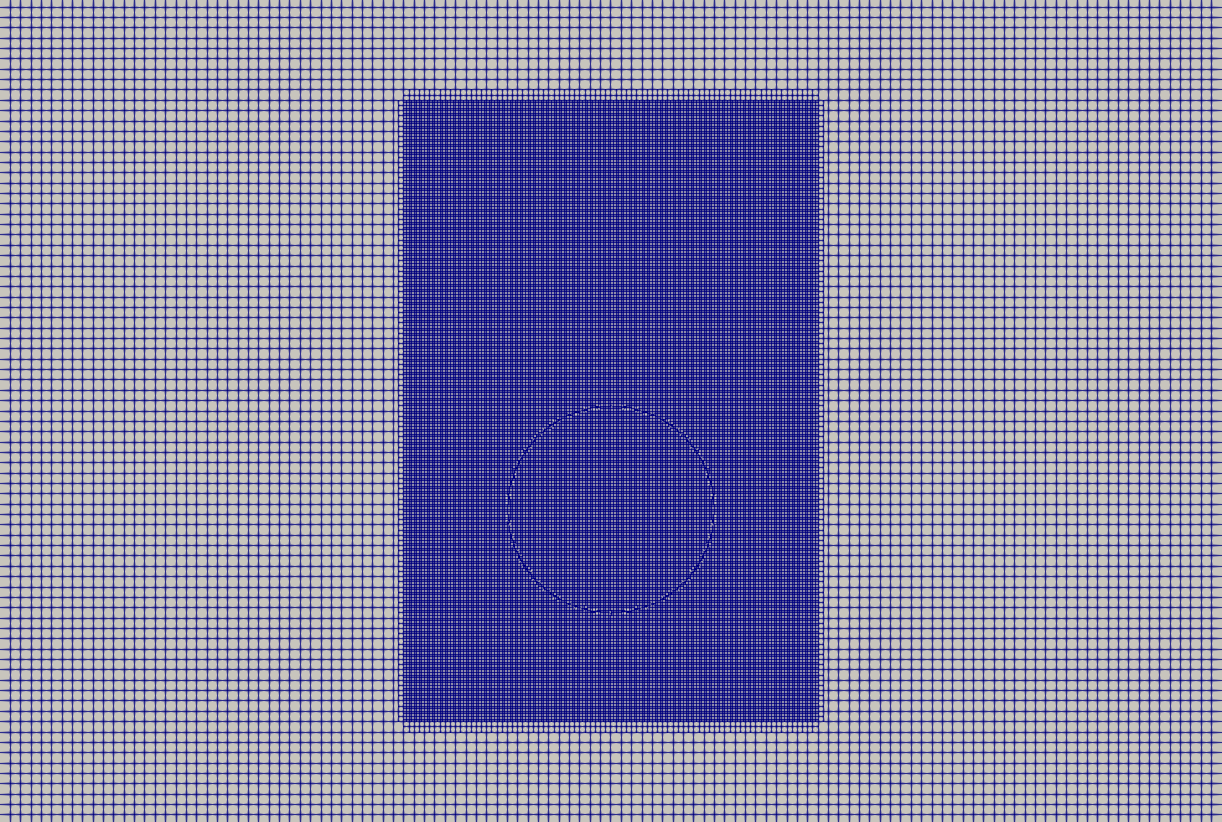}}
}
{
{\includegraphics[width=3.5cm,angle=0]{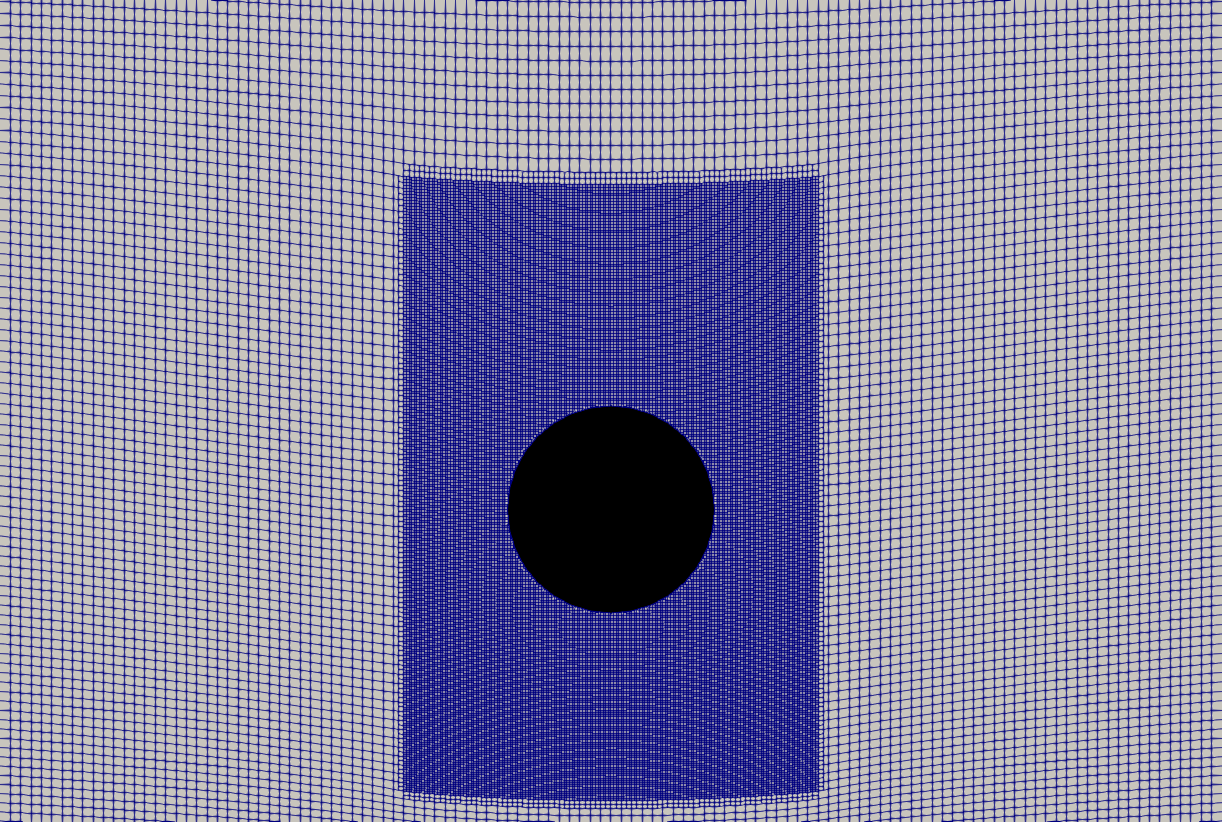}}
}
\subfigure[Mesh and velocity field for GIB ]
{
{\includegraphics[width=3.5cm,angle=0]{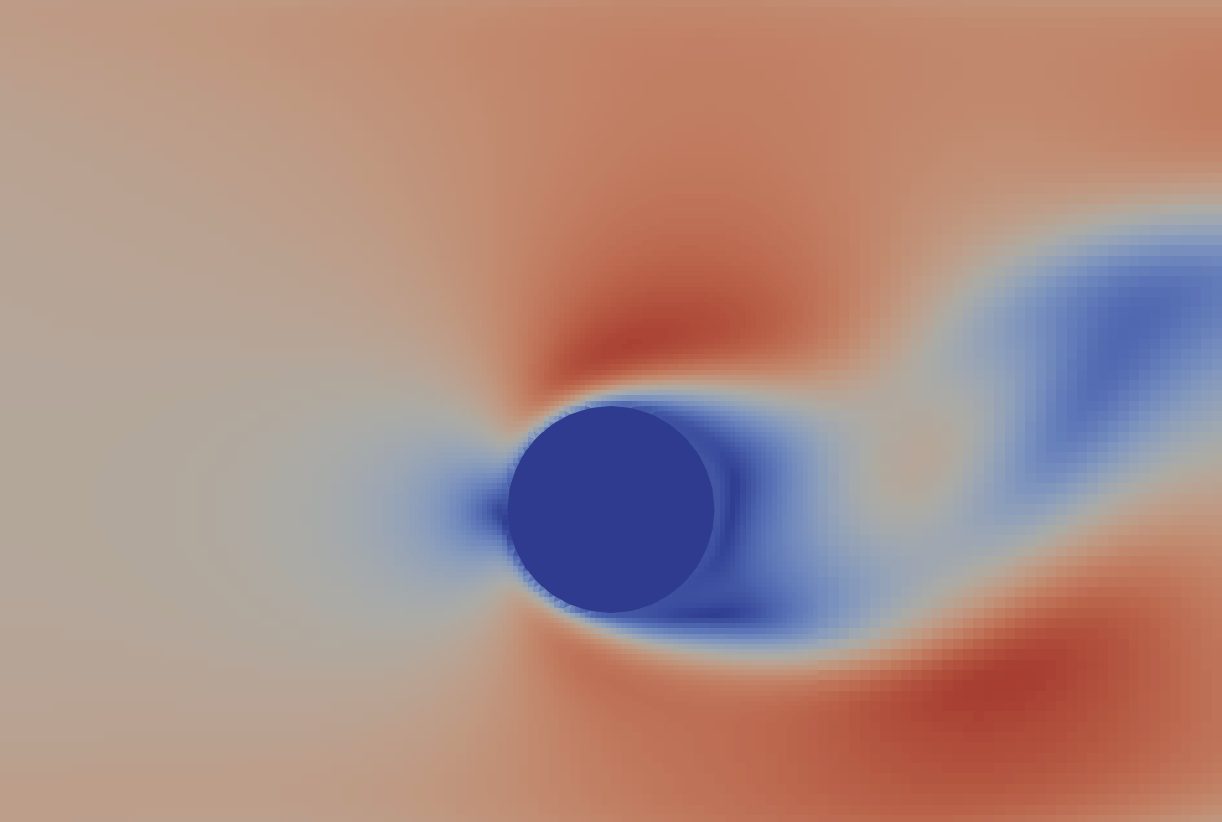}}
}
\subfigure[Mesh and Velocity field for body-fitted]
{
{\includegraphics[width=3.5cm,angle=0]{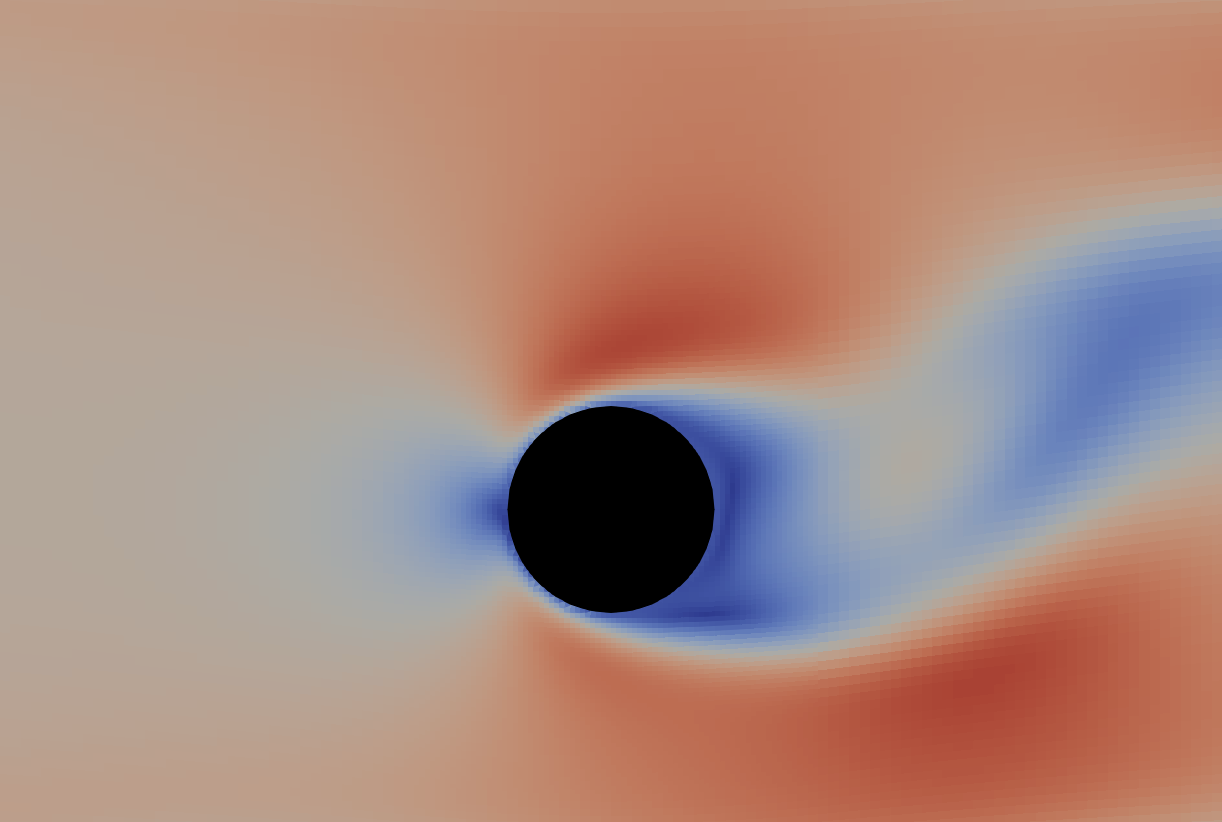}}
}
\subfigure[Mesh and velocity field for the GIB]
{
{\includegraphics[width=3.5cm,angle=0]{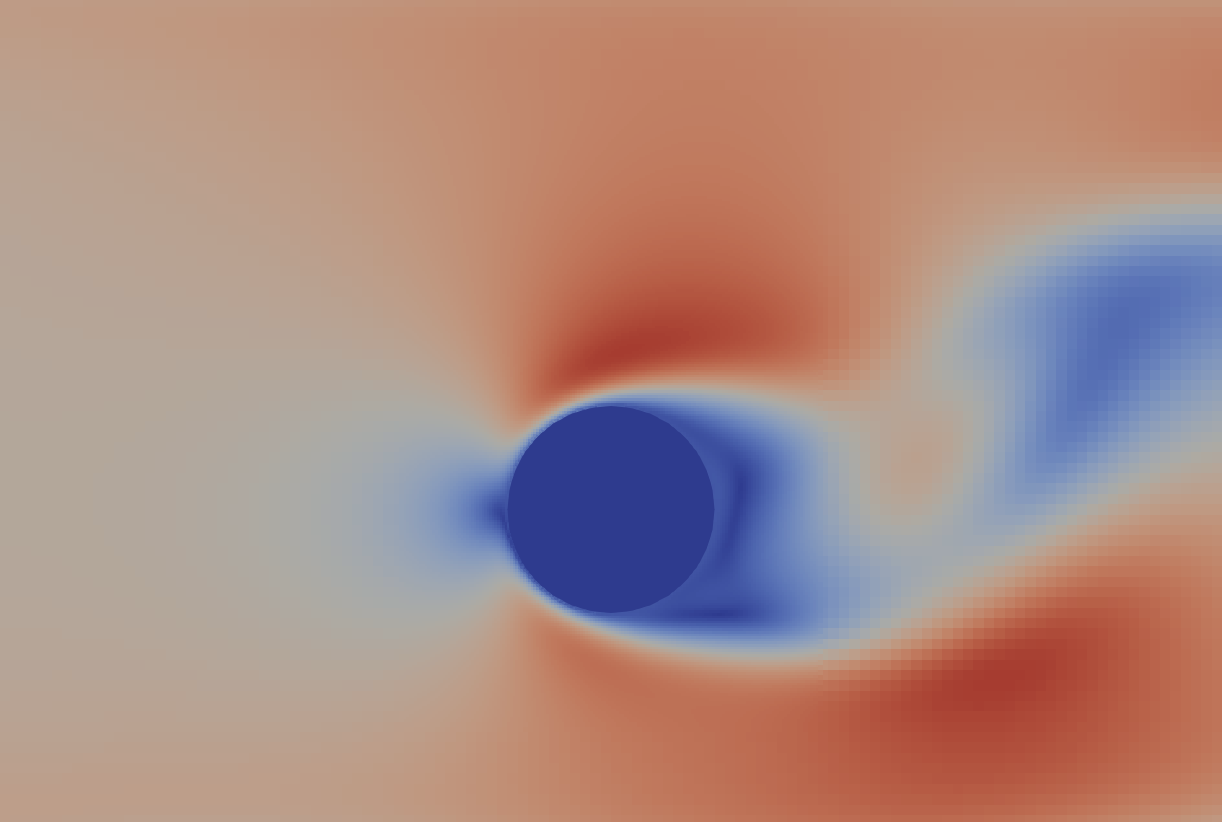}}
}
\subfigure[Mesh and velocity field for the body-fitted case]
{
{\includegraphics[width=3.5cm,angle=0]{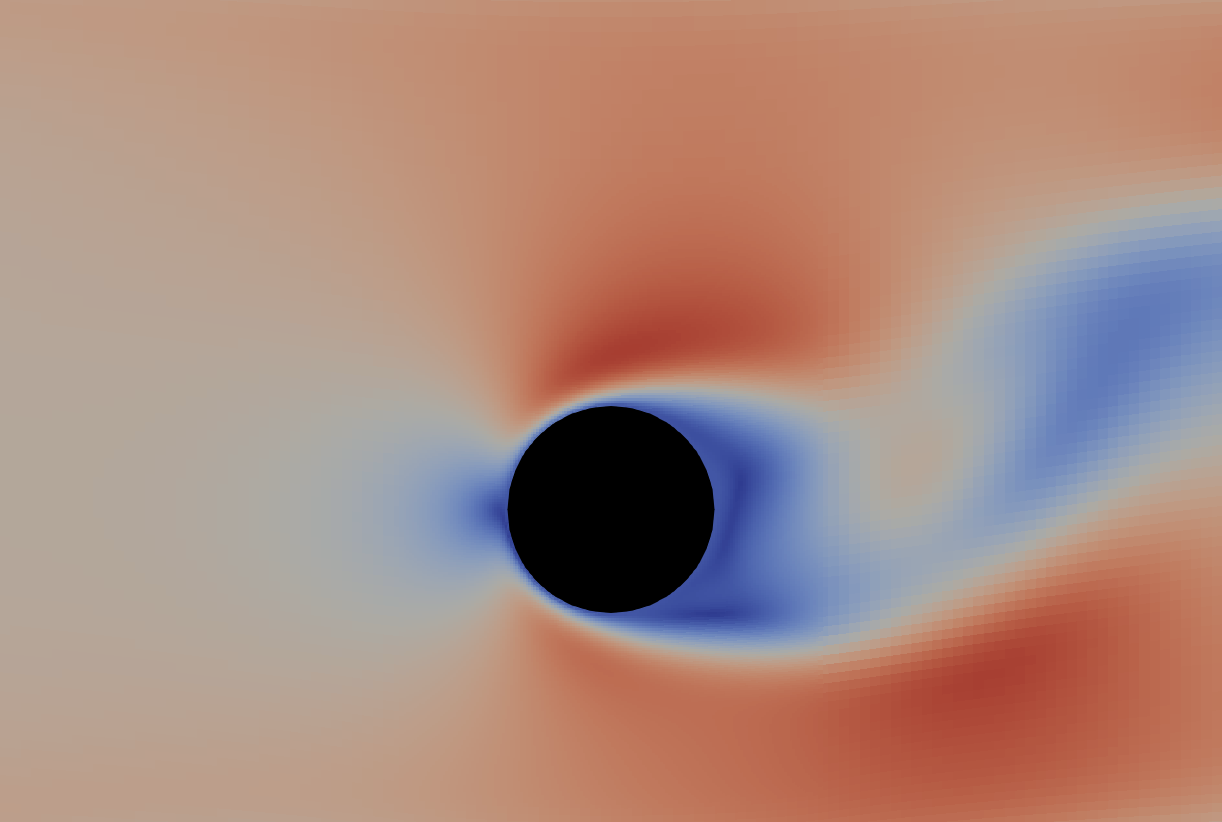}}
}
\caption {Comparison of the GIB and body-fitted case at t=20s using 1 (a,b) and 2 (c,d) levels of refinement.}
\label{val:vmoving:2lvl}
\end{figure}

\begin{figure}[h!]
\centering
\subfigure[force/time for 1 level of refinement]
{
{\includegraphics[width=7.5cm,angle=0]{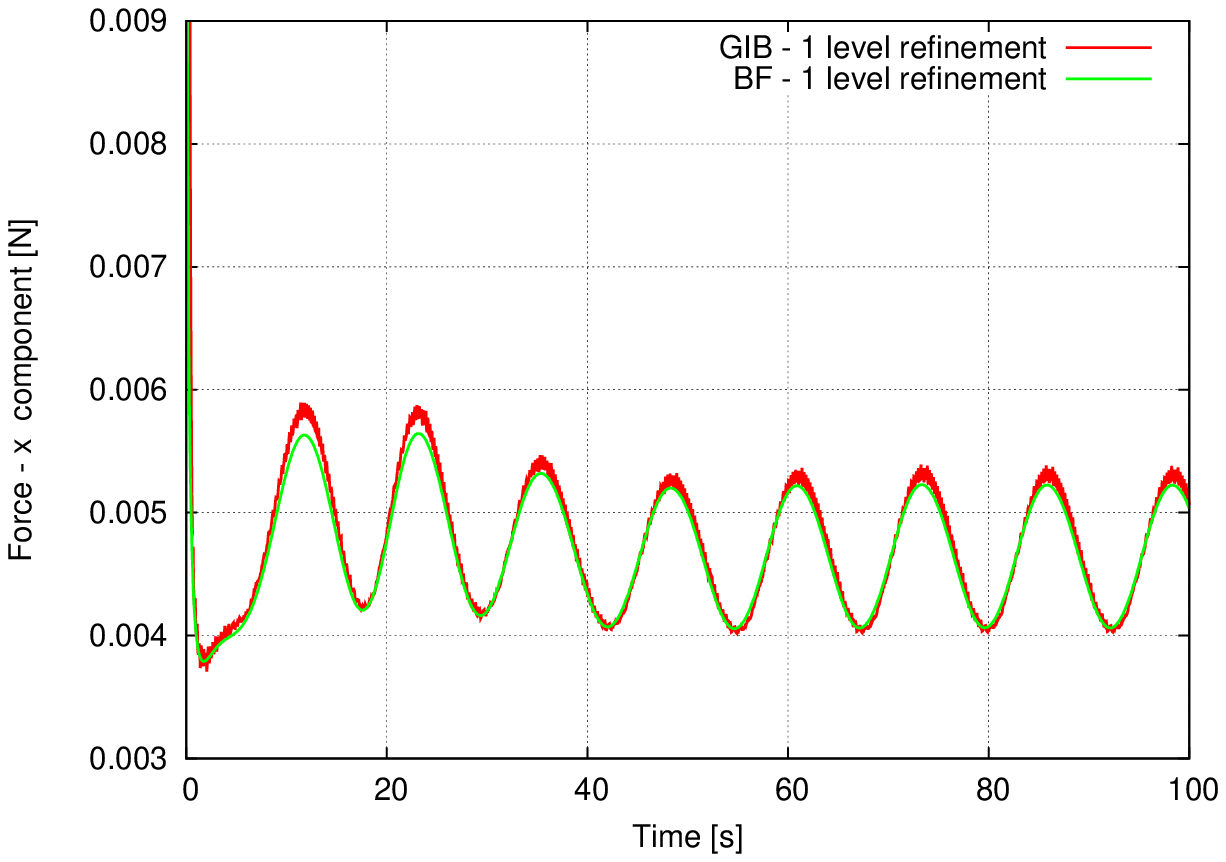}}
}
\subfigure[force/time for 2 levels of refinement]
{
{\includegraphics[width=7.5cm,angle=0]{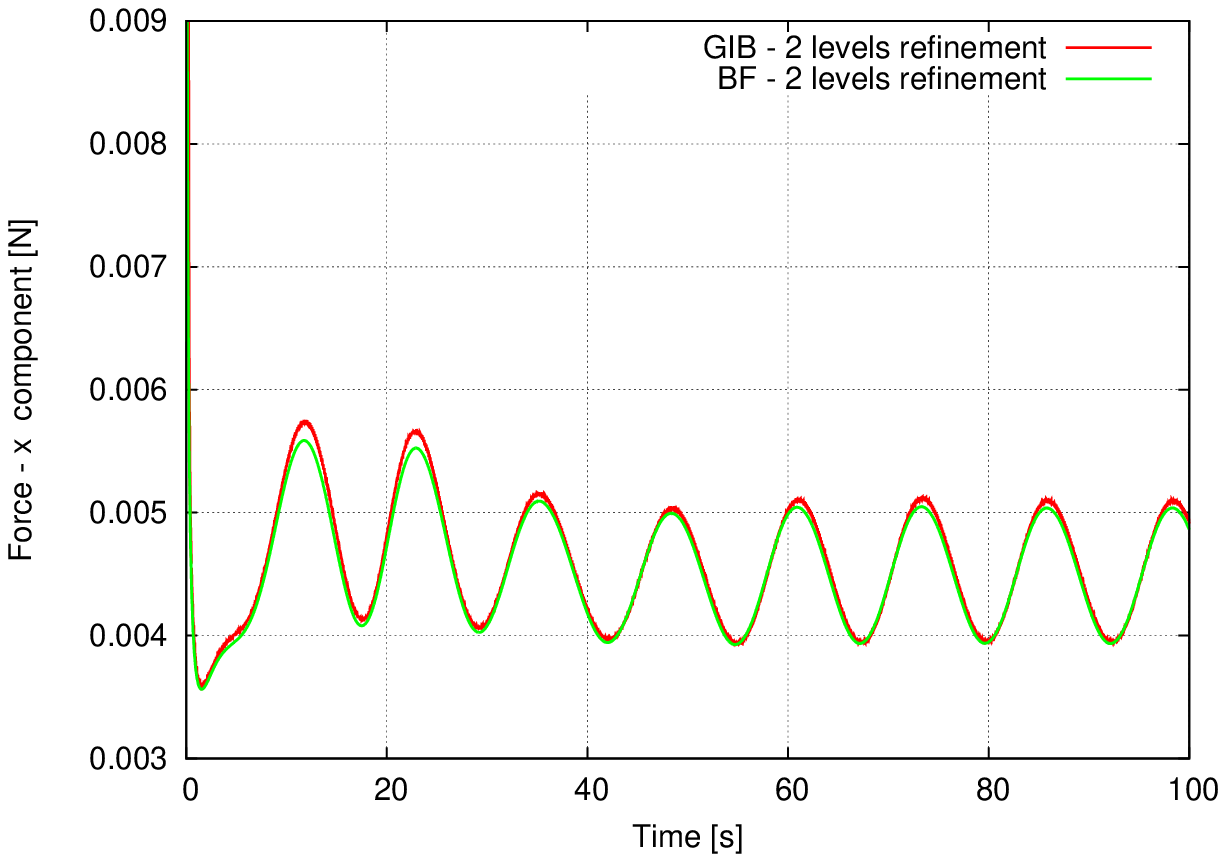}}
}
\caption{This figure represents the x component of the forces per time for 1 level (left) and 2 levels (right) of refinement.}
\label{val:vmoving:1_2forces}
\end{figure}

To further investigate this hypothesis, we successively refine the mesh in the vicinity of the vertically oscillating cylinder. In figure \ref{val:vmoving:2lvl} the GIB case is compared with the body-fitted case using both one and two levels of refinement relative to the initial case. As before, the velocity fields in the GIB and body-fitted cases appear superficially very similar. It is also clear from visual inspection of the force graphs (figure \ref{val:vmoving:1_2forces}) that the reduction in mesh size is very effective in reducing the spurious oscillations in the integral signal. In table \ref{force_dif}, the root mean square of the difference between the force integral from the GIB and the body fitted results (equation \ref{DForce}) is represented for each resolution. A clear trend of reduction in the error signal with reduced mesh spacing is visible. 

\begin{eqnarray}
        \Delta F_{rms} = \sqrt
        {
            \frac{\sum_{i=1}^{n}{{(force_{x}}_{GIB}-{force_{x}}_{BF})^2}}{n}
        }
		\label{DForce} 
\end{eqnarray}

\begin{table}[h]
  \begin{center}
    \begin{tabular}{| c | c | c |}
    \hline
    Case   &  cell size($m^3$) &   $\Delta F_{rms}$    \\
    \hline  \hline
    1 & 1e-4  & 1.16e-04    \\
    \hline
    2 & 1e-6  & 0.686e-04    \\
    \hline 
    3 & 1e-8  & 0.508e-04    \\
    \hline 
    \end{tabular}
  \end{center}
\caption{Force comparison of the GIB and the Body-fitted mesh for different base mesh sizes.}
\label{force_dif}
\end{table}

\vspace{60mm}

\subsection{Butterfly Valve}\label{valve}
\begin{figure}[h!]
  \begin{center}
    \includegraphics[width=7.5cm,angle=0]{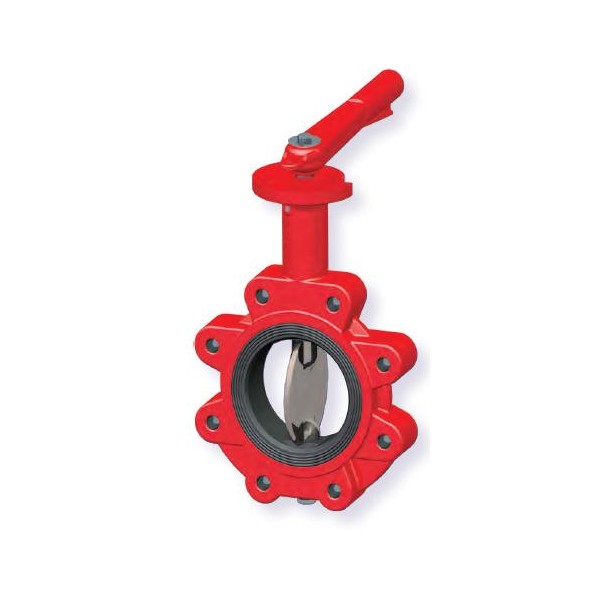}
    \label{valve:real}
    \caption{\textbf{Weco\textregistered} 22L model Butterfly valve}
  \end{center}
\end{figure}

In this section, the proposed approach is validated against experimental results for a butterfly valve. The butterfly valve is a \textbf{Weco\textregistered} 22L model from \textbf{FMCTechnologies} \cite{valve} and it is represented in figure \ref{valve:real}.

The experimental results in terms of mass flow and pressure drop are provided from the manufacturer. With a given pressure drop, the mass flow is calculated using the following expression:

\begin{eqnarray}
        Q = C_v \sqrt{ \frac{\Delta P}{G} }
		\label{massFlow_exp} 
\end{eqnarray}

where $Q[gpm]$ is the mass flow, $\Delta P[psi]$ is the pressure drop, the $C_v$ is the valve constant which is provided from the manufacturer. For a 20 inch valve the $C_v$ is shown for different positions of the valve in the table \ref{constantCv}. Constant $G$ is the specific gravity of fluid and its value for water is 1.

\begin{table}[h]
  \begin{center}
    \begin{tabular}{| c | c | c | c | c | c | c | c | c | c |}
    \hline
    Valve Size[in.]  &  $10^o$ & $20^o$ & $30^o$ & $40^o$ & $50^o$ & $60^o$ & $70^o$ & $80^o$  &  $90^o$    \\
    \hline  \hline
    20 & 161 & 723 & 1500 & 2700 & 4800 & 7900 & 12500 & 18500 & 27000 \\
    \hline 
    \end{tabular}
  \end{center}
\caption{$C_v$ value for various disc angles of 20 inch size valve.}
\label{constantCv}
\end{table}

\begin{figure}[h!]
\centering
\subfigure[Valve position inside the pipe]
{
{\includegraphics[width=7.5cm,angle=0]{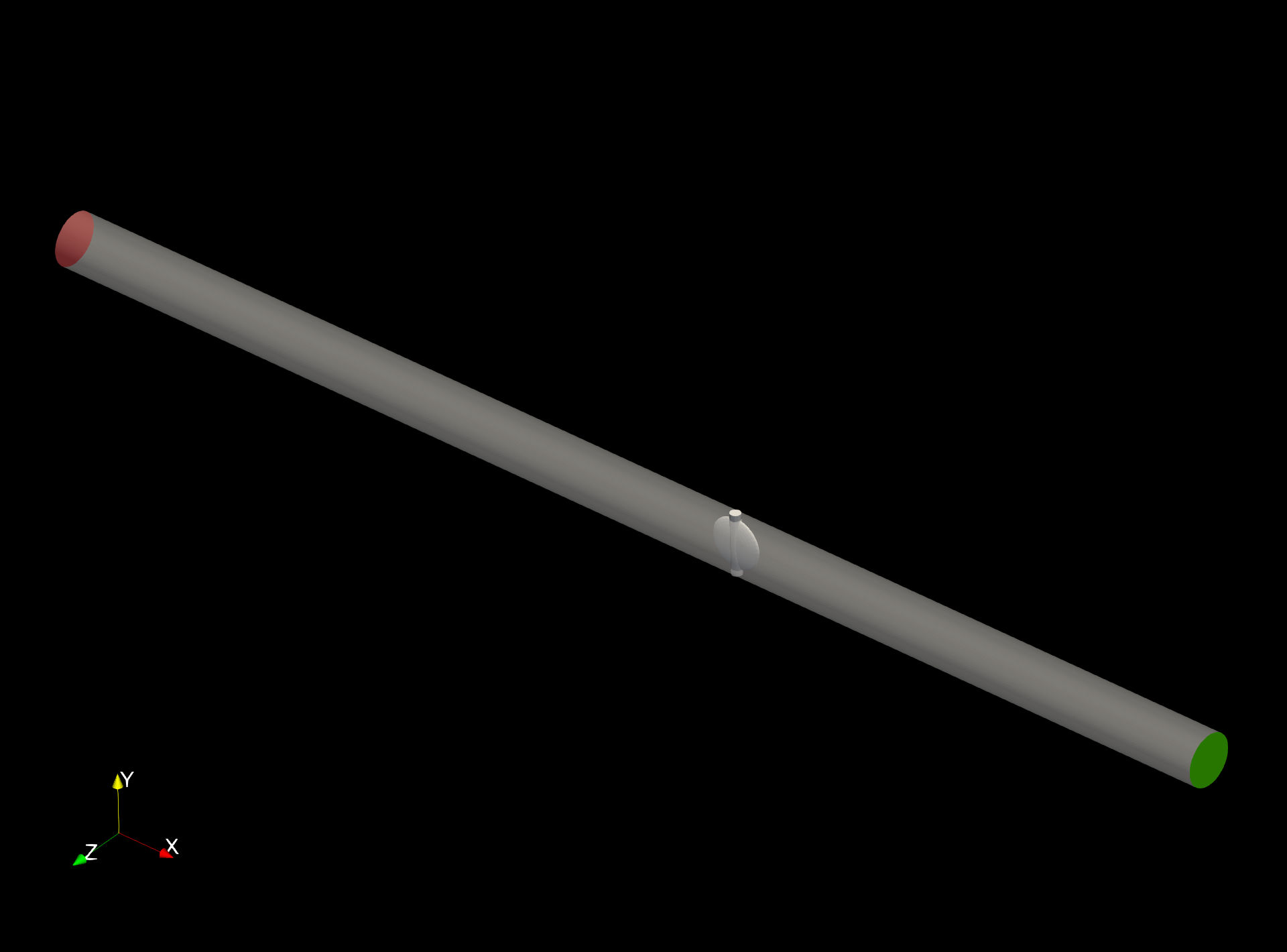}}
}
\subfigure[Volume refinements close to the valve]
{
{\includegraphics[width=7.5cm,angle=0]{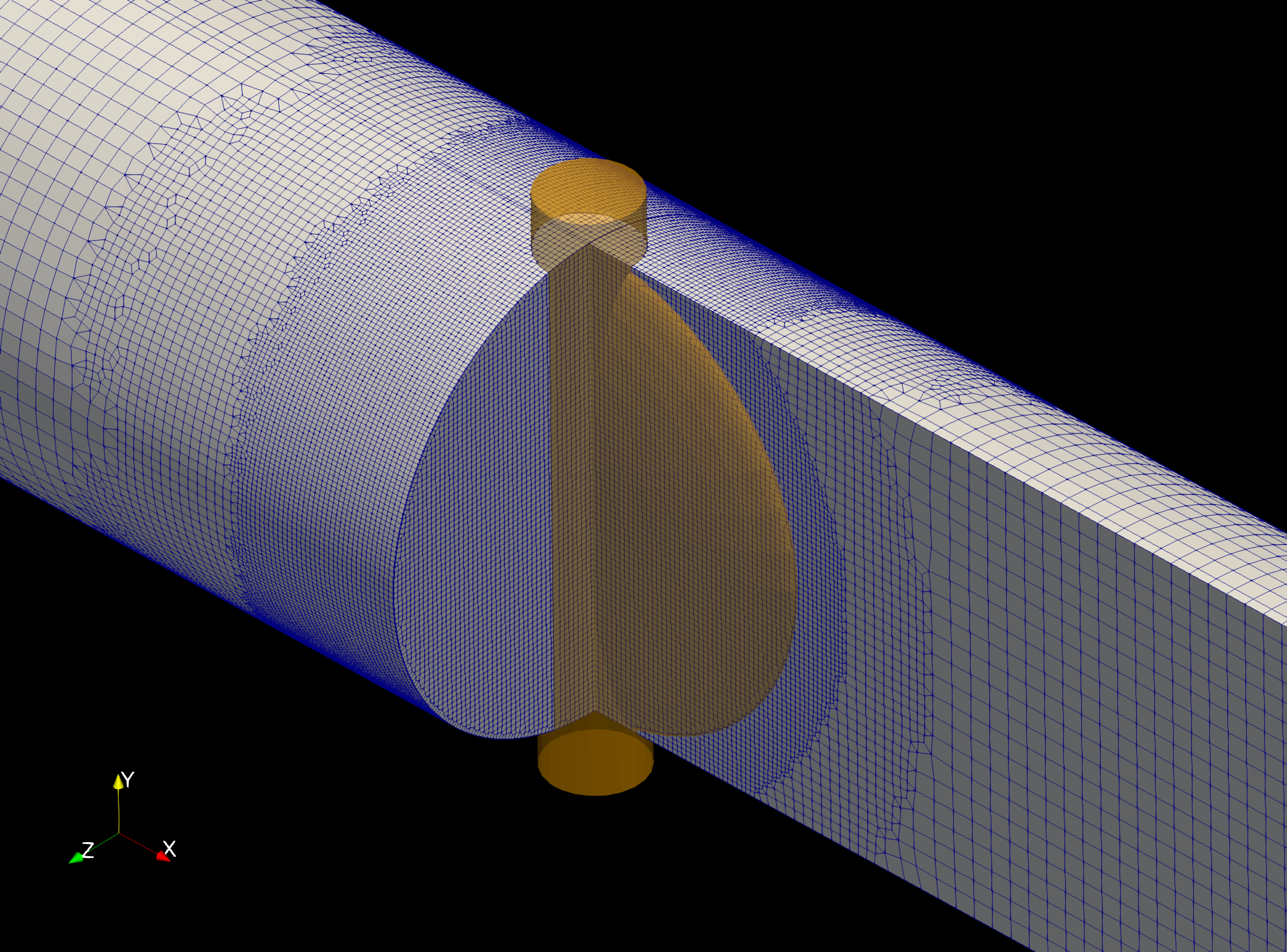}}
}
\caption{Computational mesh}
\label{valveMesh}
\end{figure}

The computational mesh has 1 million cells and it is represented in figure \ref{valveMesh}. The inlet (green) and the outlet (red) patch are located 10 and 14 diameters away from the valve, respectively. The fluid is water with density $\rho = 998.2 [kg/m^3]$ and kinematic viscosity $\nu = 1.0038*10^{-6} [m^2/s]$. 
Total pressure boundary condition is specified in the inlet $p_{tot} = 728.54 [Pa]$. The outlet static pressure is fixed to the value of 0. The $kOmega-SST$ is chosen for turbulence model and wall functions are applied to the wall boundaries (including the GIB boundaries).

\begin{figure}[h!]
\centering
\subfigure[Static GIB (red) vs experimental results (green)]
{
{\includegraphics[width=7.5cm,angle=0]{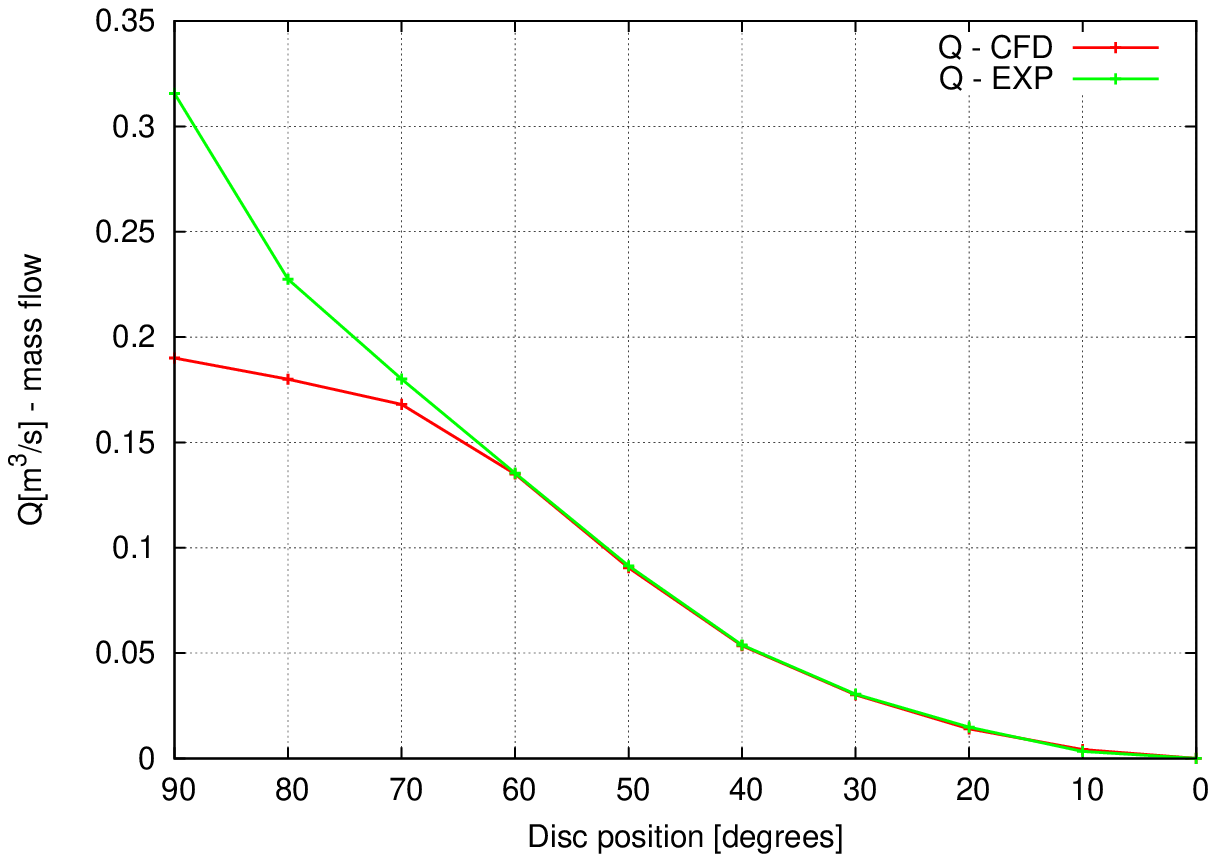}}
\label{valve:stativvsxp}
}
\subfigure[Static GIB (green) vs moving GIB (red)]
{
{\includegraphics[width=7.5cm,angle=0]{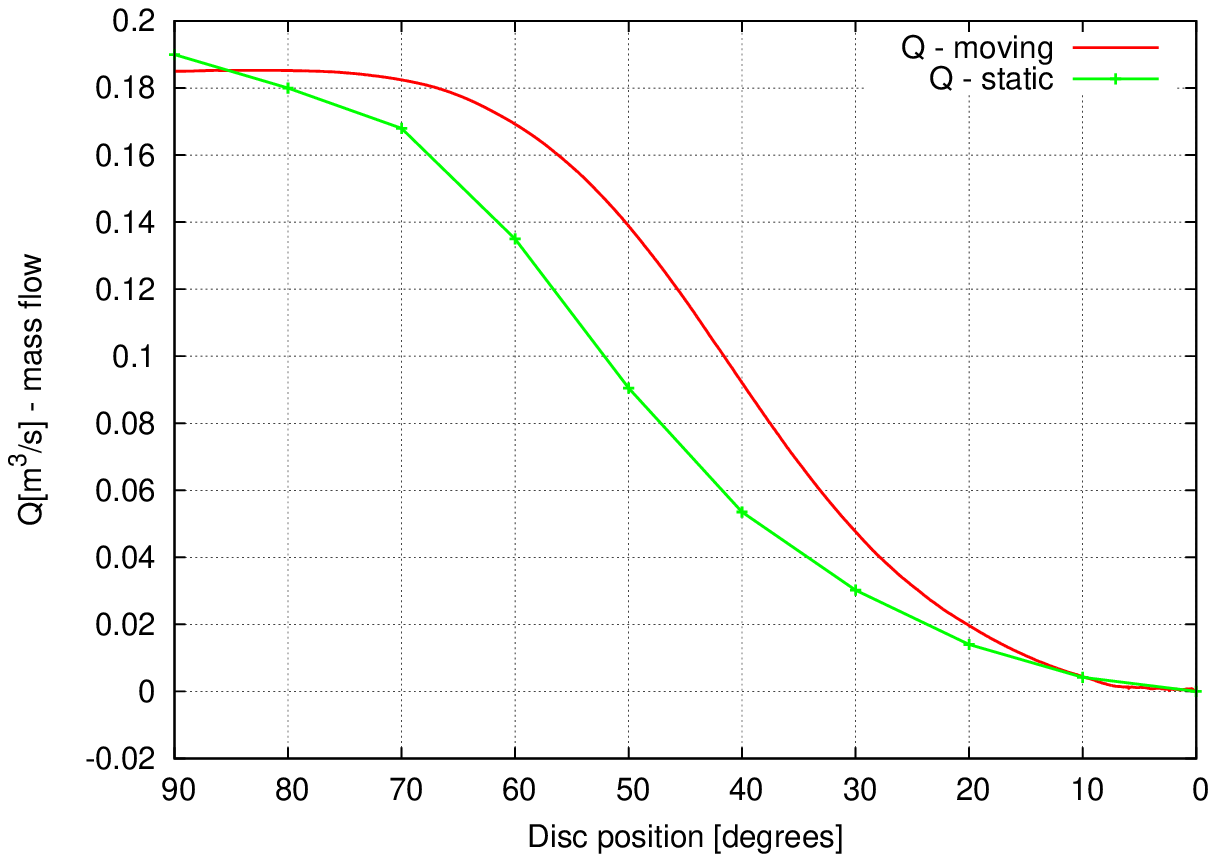}}
\label{valve:stativvsmoving}
}
\caption {Left: Comparison between the predicted mass flow of static GIB in different positions of the disc and the experiments. Right: Mass flow predicted from static and moving GIB}
\label{valve:results}
\end{figure}

In the first stage, the disc is modelled using static GIB for different positions (per $10^o$ degrees) and the results are compared against the experimental results (figure  \ref{valve:stativvsxp}). The mass flow $Q$ predicted from the simulation and the experiments in most positions have a good agreement. The CFD fails to calculate accurately the mass flow only when the valve is wide open. However, this can be due to turbulent boundary conditions or difference in the position where the the pressure drop is measured in the experiments and the simulation.

%
%

\begin{figure}[h!]
\centering
\subfigure[U - $\phi$=90 deg]
{
{\includegraphics[width=3.75cm,angle=0]{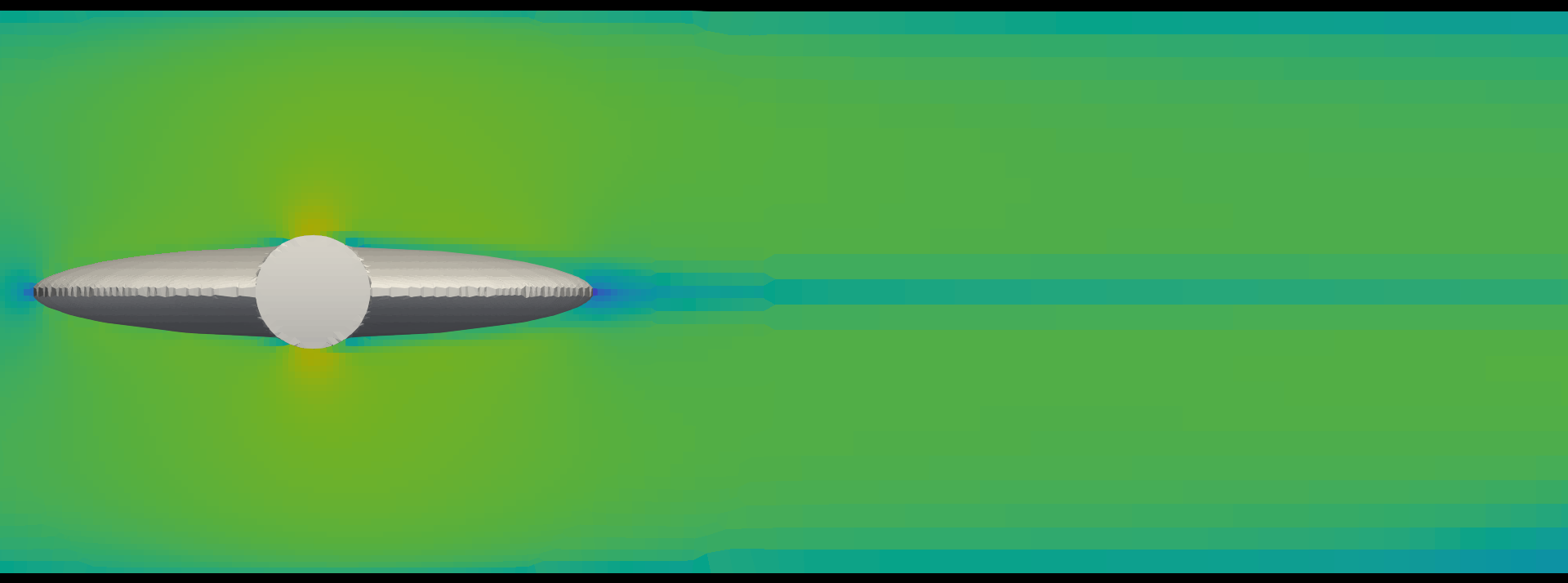}}
\label{valve:U:90}
}
\subfigure[U - $\phi$=60 deg]
{
{\includegraphics[width=3.75cm,angle=0]{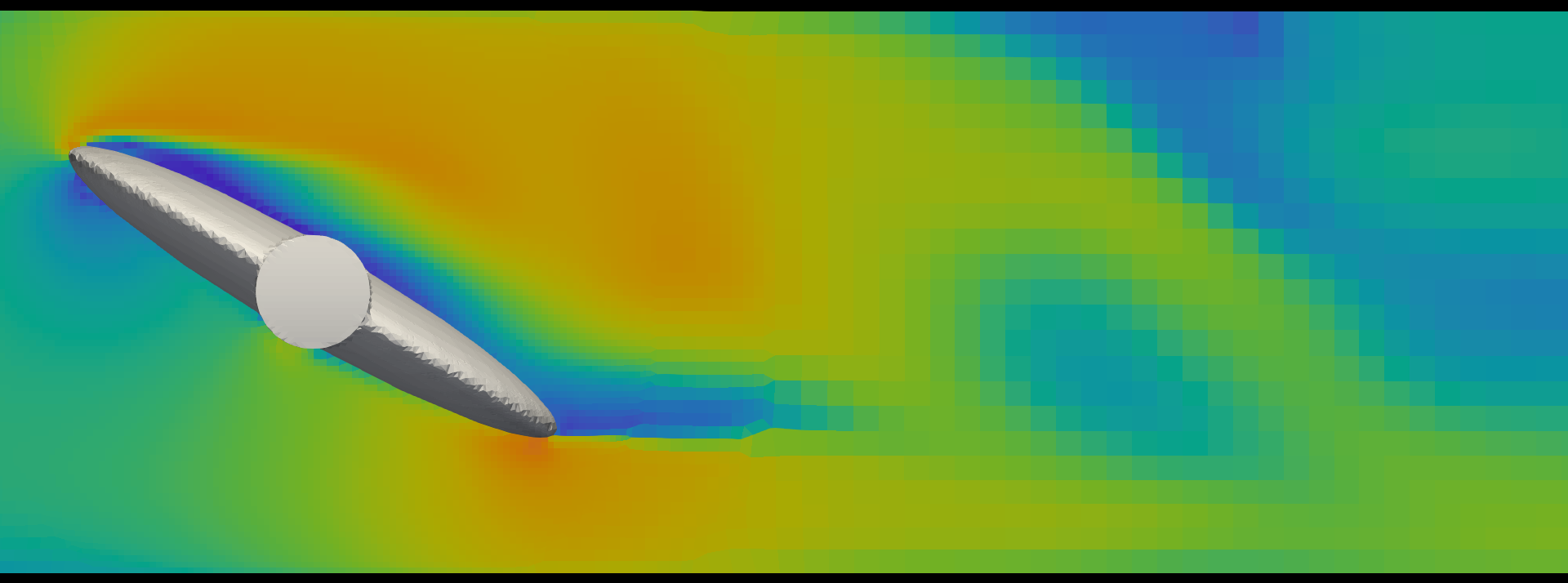}}
\label{valve:U:60}
}
\subfigure[U - $\phi$=30 deg]
{
{\includegraphics[width=3.75cm,angle=0]{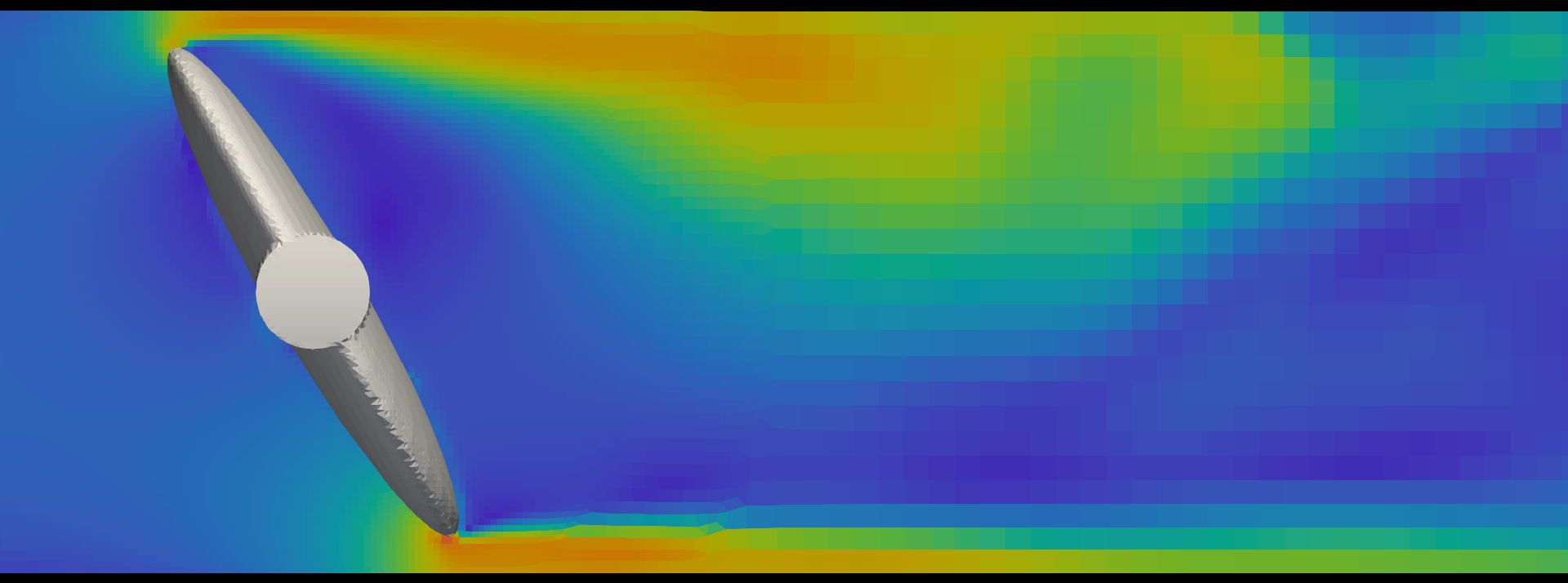}}
\label{valve:U:30}
}
\subfigure[U - $\phi$=0 deg]
{
{\includegraphics[width=3.75cm,angle=0]{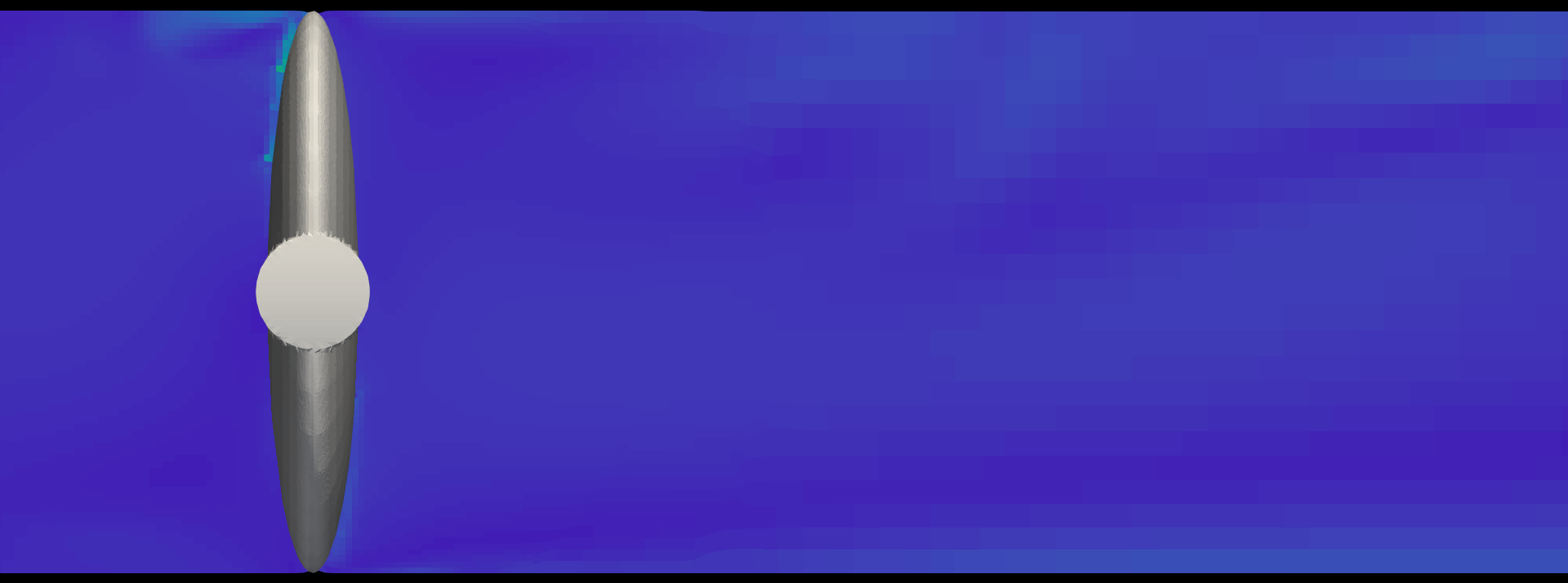}}
\label{valve:U:0}
}
\subfigure[p - $\phi$=90 deg]
{
{\includegraphics[width=3.75cm,angle=0]{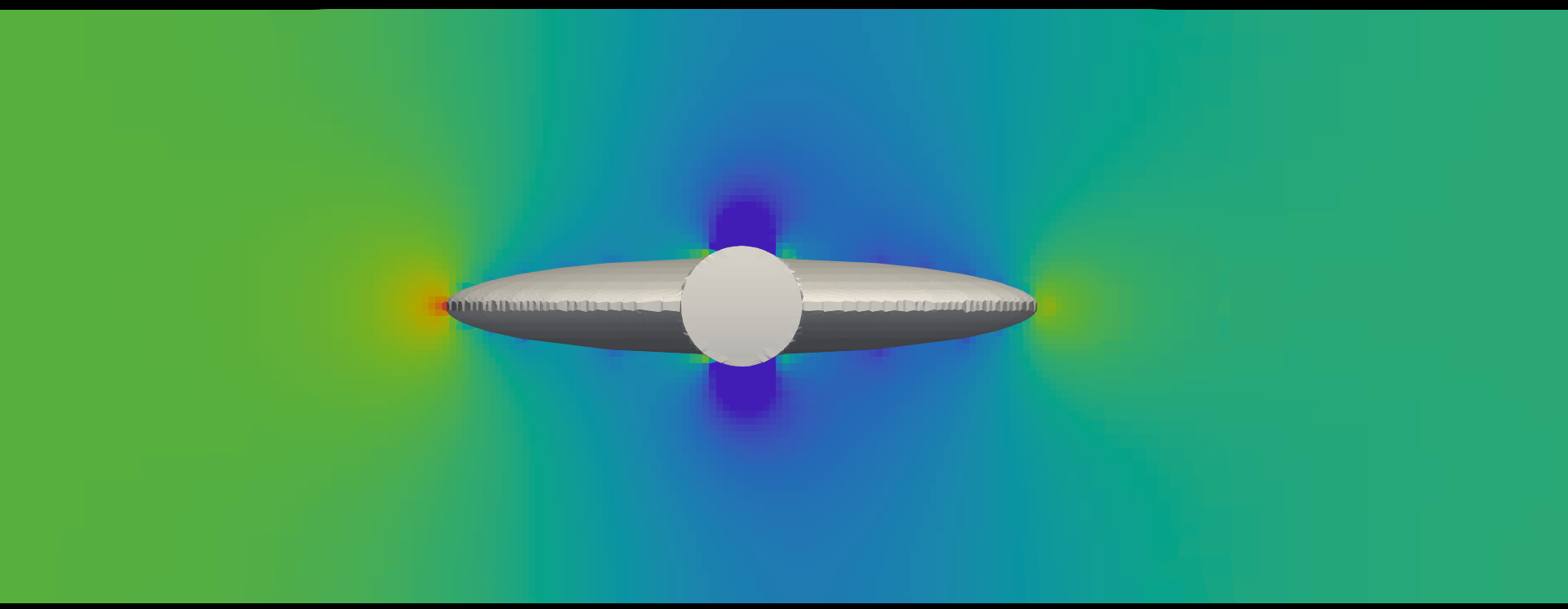}}
\label{valve:p:90}
}
\subfigure[p - $\phi$=60 deg]
{
{\includegraphics[width=3.75cm,angle=0]{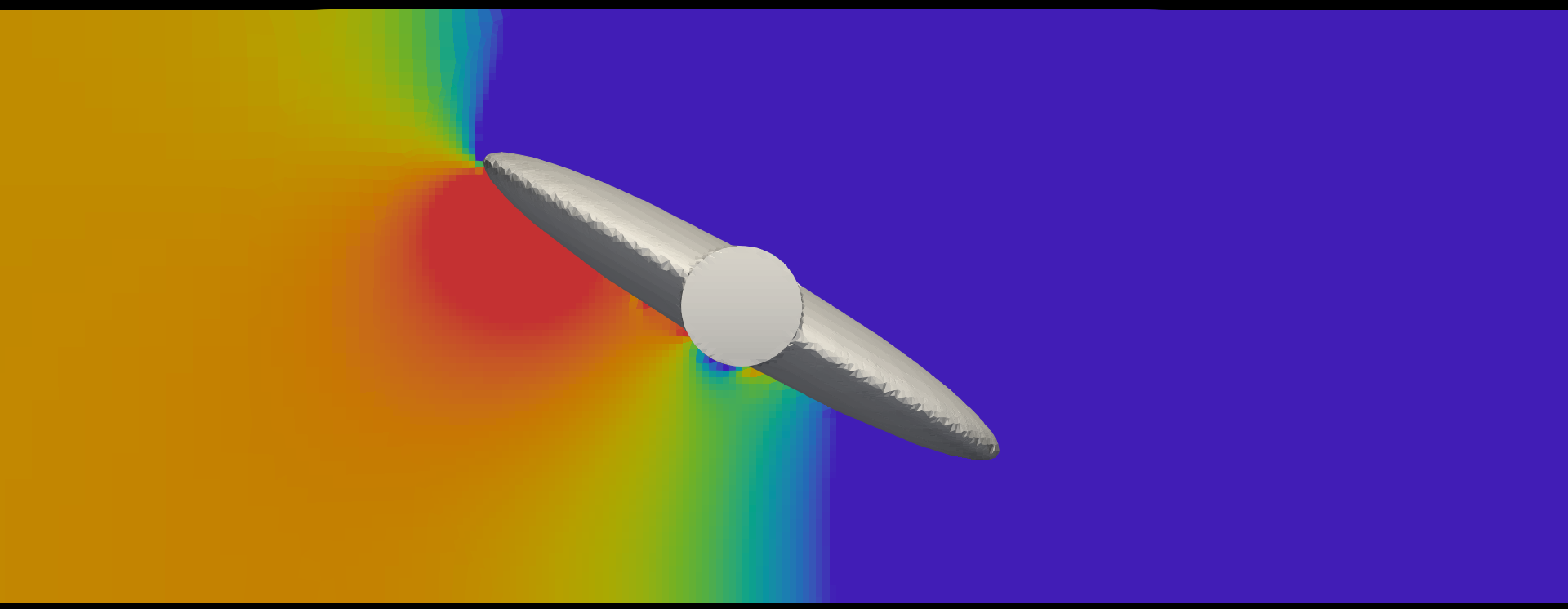}}
\label{valve:p:60}
}
\subfigure[p - $\phi$=30 deg]
{
{\includegraphics[width=3.75cm,angle=0]{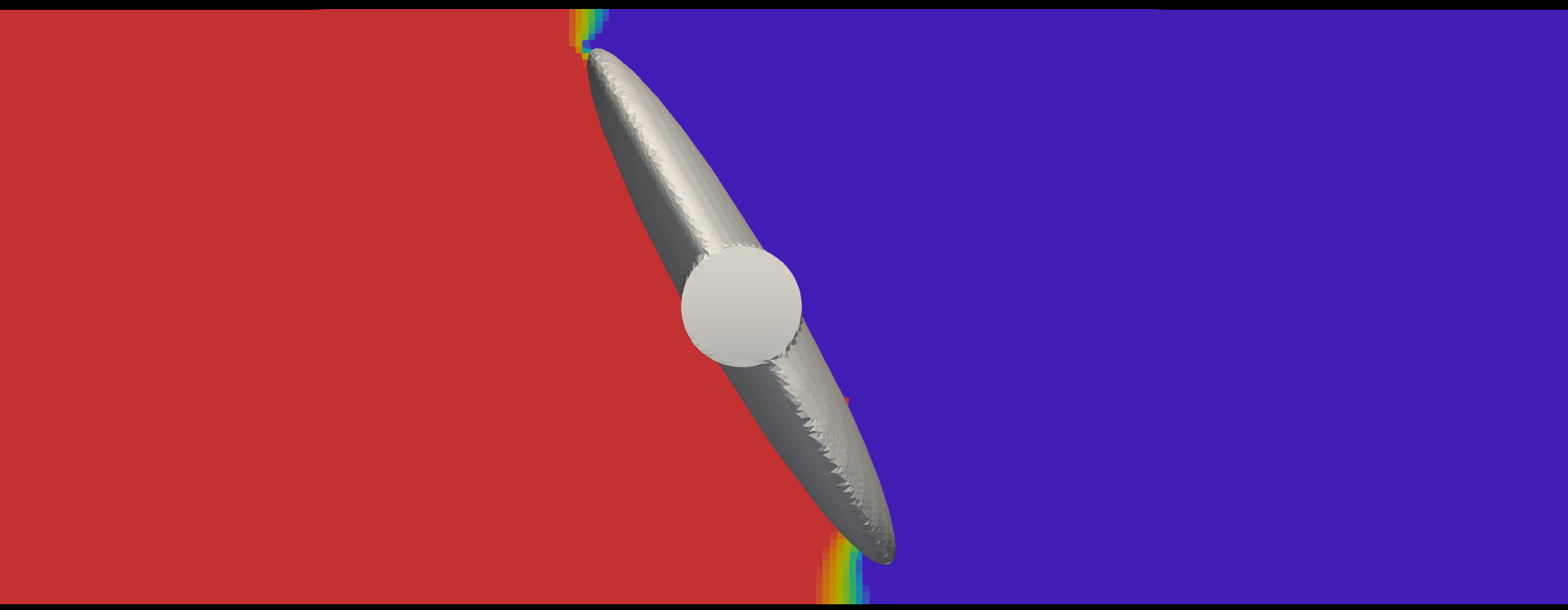}}
\label{valve:p:30}
}
\subfigure[p - $\phi$=0 deg]
{
{\includegraphics[width=3.75cm,angle=0]{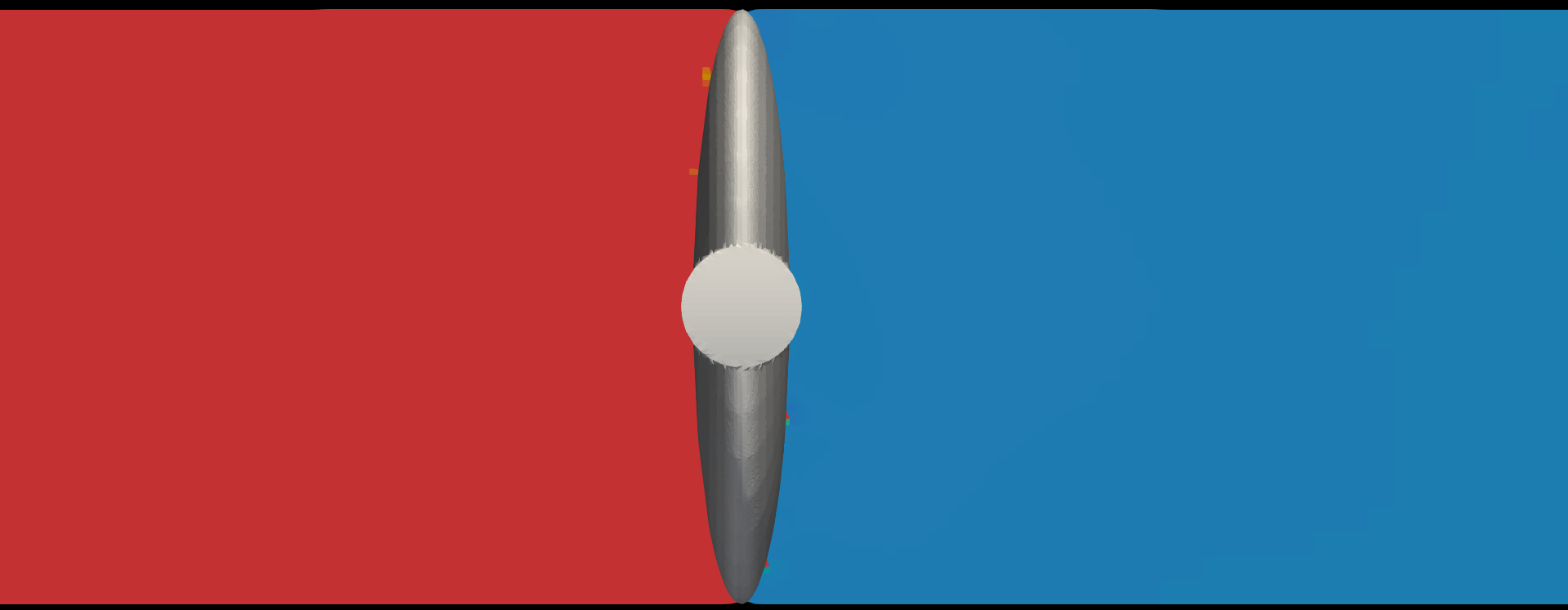}}
\label{valve:p:0}
}
\caption {Velocity (top) and pressure (bottom) fields for different positions of the disc}
\label{valve:Up}
\end{figure}

\begin{figure}[h!]
\centering
\subfigure[$\nu_t$ - Turbulent kinematic viscosity]
{
{\includegraphics[width=5cm,angle=0]{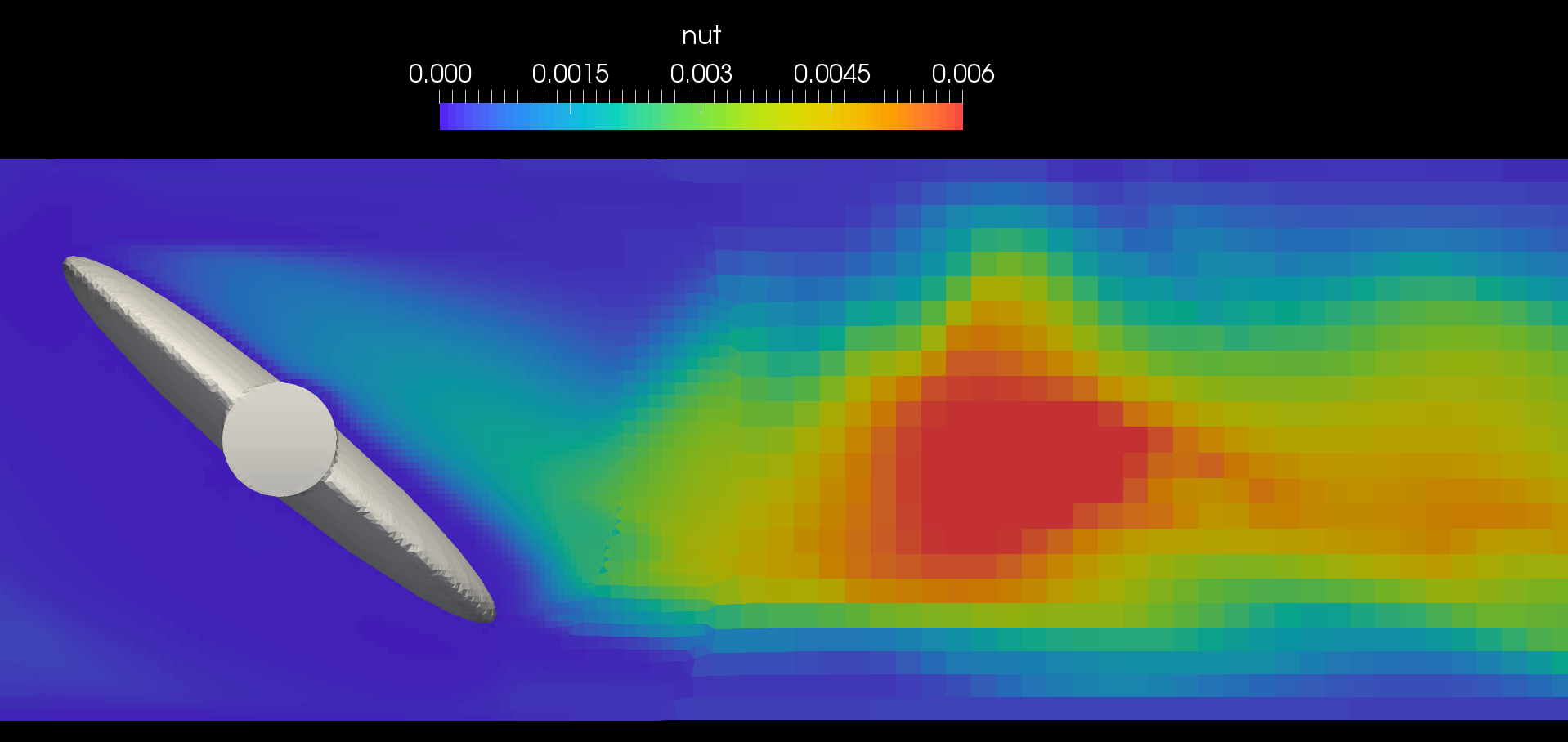}}
\label{valve:nut:90}
}
\subfigure[$\kappa$ - Turbulent kinetic energy]
{
{\includegraphics[width=5cm,angle=0]{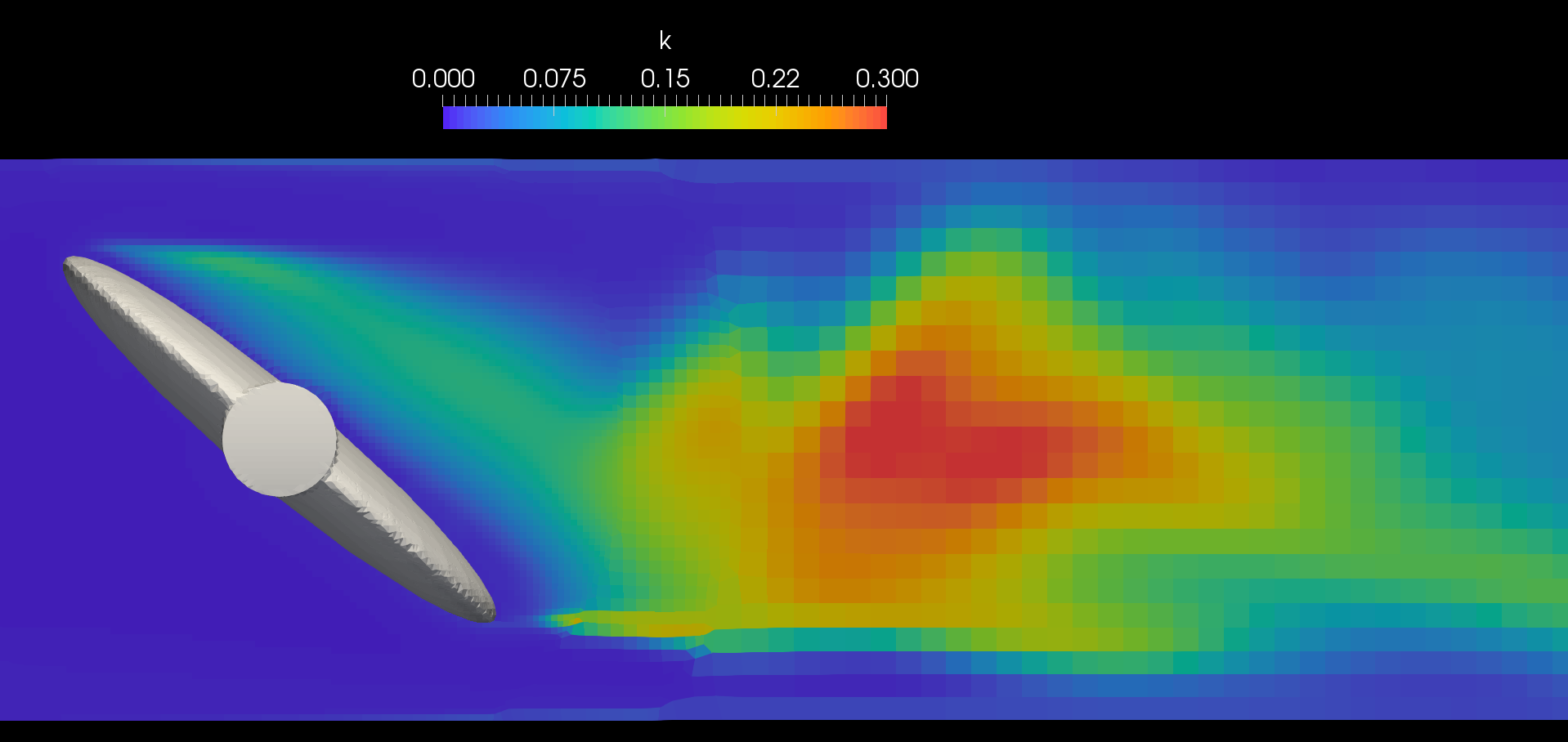}}
\label{valve:k:60}
}
\subfigure[$\omega$ - Specific dissipation]
{
{\includegraphics[width=5cm,angle=0]{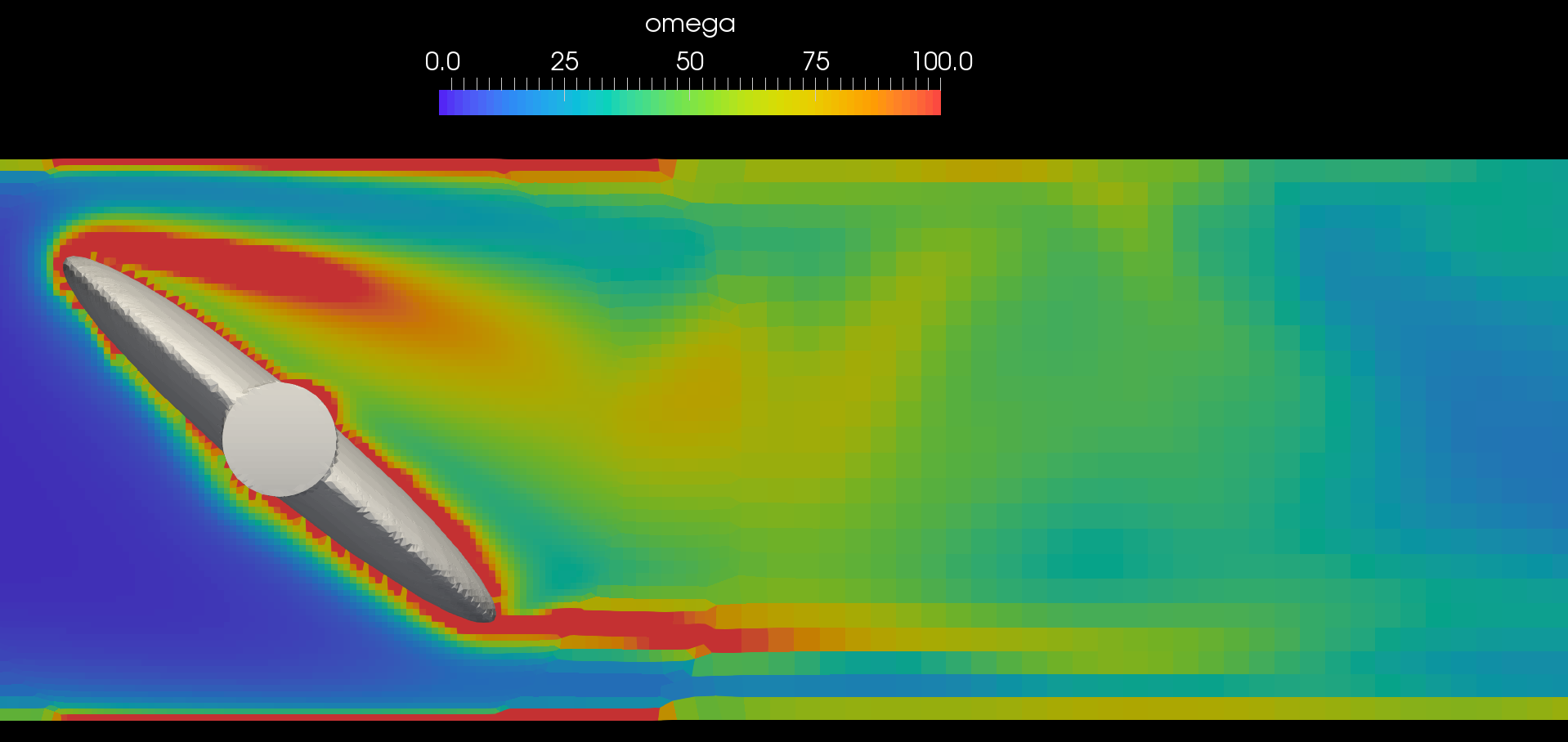}}
\label{valve:omega:30}
}
\caption {Turbulent variables when the disc is at position $\phi$=50 deg}
\label{valve:turb}
\end{figure}

In the second stage, rotating motion of  $\omega = 0.087$ $rad/s$ ($5^o$ degrees/sec) is added to the disc for modelling a closing valve. In the beginning, a steady state run is performed for an open valve to get a fully developed flow. After, the valve starts to close, until the flow is completely blocked after 18 sec. The velocity and pressure fields for different positions of the disc are represented in figure \ref{valve:Up}. Regarding turbulence, $\kappa$ (turbulent kinetic energy), $\omega$ (specific dissipation) and $\nu_t$ (turblent kinematic viscosity) are presented in figure \ref{valve:turb} for disc position $\phi=50$ degree.

In figure \ref{valve:stativvsmoving} is shown the comparison between static disc simulations and a moving disc for mass flow. There is a difference due to the unsteady effects of the moving valve.

\section{CONCLUSIONS}

In this paper, we presented a new method for rigid body motion in the finite volume context. It is shown that the GIB provides the same level of accuracy as a traditional peripheral boundaries, but can be placed anywhere with the original domain and can support arbitrarily large motions of the surface. 

One of the big advantages of the GIB approach is that it can be added to existing finite volume codes with minimal changes in structure. This stems from the fact that the GIB matrix contributions can be formulated in a supplementary form which can be added to every finite volume operation. Unlike the cut-cell method, the GIB does not introduce any additional computational elements into the mesh. This, coupled to the supplemental form of the boundary contribution, means that no changes to the addressing structure of the grid is necessary. Although off-diagonal matrix coefficients might become zero, their entries remain. This has significant advantages in terms of performance, since, unless the algorithm was specifically designed to maximise dynamic addressing performance, recalculation of mesh addressing will constitute considerable computational overheads in dynamic mesh calculations. In addition, the fact that the GIB is a normal boundary coupled to local mesh deformation, means that existing routines for boundary specification should be reusable in any object oriented context.

We should note that our initial implementation of the moving GIB boundary is limited to first order accuracy in time. Extension to second order will be undertaken as part of continued work on the subject and should be straight-forward. Furthermore, the accuracy of flow in the vicinity of the GIB boundary suffers from the lack of anisotropic layer refinement, just like cut-cell and other IB methods. We are however exploring the possibility of using the GIB as an "of-the-wall" grid interface, which should alleviate at least some of these concerns (at the cost of additional complexity). Finally, the most challenging aspect of the GIB method is aligning the internal facets of the mesh with the surface without creating invalid elements. This step becomes more difficult with increased geometric complexity, but has to be kept as efficient as possible to prevent it from dominating the computational cost of the algorithm.

While the GIB approach has proven its capability in the modelling of rigid-body motion, it is also showing great promise in the modelling of more plastic deformations. To date, we have been able to extend the GIB framework to model solid-combustion and adjoint-based topology/shape optimization 3D applications, where the interface motion depends on the solution of fluid properties. The method can also efficiently handle topological changes and coupled fluid-solid interactions, possibilities which we look forward to exploring in future.

\section{ACKNOWLEDGEMENTS}

The present work has been conducted under the auspices of the ITN Aboutflow FP7 EU project \cite{aboutFlow}.

\section*{References}

\bibliography{GIB_arxiv_nov}

\end{document}

%% file: macrosKarpou.tex
\def\pp{\partial}

%% file: GIB_arxiv_nov.bbl
\begin{thebibliography}{10}
\expandafter\ifx\csname url\endcsname\relax
  \def\url#1{\texttt{#1}}\fi
\expandafter\ifx\csname urlprefix\endcsname\relax\def\urlprefix{URL }\fi
\expandafter\ifx\csname href\endcsname\relax
  \def\href#1#2{#2} \def\path#1{#1}\fi

\bibitem{Peskin1972}
C.~Peskin, Flow patterns around heart valves: A numerical method, Journal of
  Computational Physics 10~(2) (1972) 252--271.
\newblock \href {http://dx.doi.org/10.1016/0021-9991(72)90065-4}
  {\path{doi:10.1016/0021-9991(72)90065-4}}.

\bibitem{Peskin1977}
C.~S. Peskin,
  \href{http://www.sciencedirect.com/science/article/pii/0021999177901000}{Numerical
  analysis of blood flow in the heart}, Journal of Computational Physics 25~(3)
  (1977) 220 -- 252.
\newblock \href
  {http://dx.doi.org/http://dx.doi.org/10.1016/0021-9991(77)90100-0}
  {\path{doi:http://dx.doi.org/10.1016/0021-9991(77)90100-0}}.
\newline\urlprefix\url{http://www.sciencedirect.com/science/article/pii/0021999177901000}

\bibitem{Peskin2002}
C.~S. Peskin, The immersed boundary method, Acta Numerica 11 (2002) 479–517.
\newblock \href {http://dx.doi.org/10.1017/S0962492902000077}
  {\path{doi:10.1017/S0962492902000077}}.

\bibitem{mohdyusof1997}
J.~Mohd-Yusof, {Combined Immersed-Boundary/B-spline methods for simulations of
  flow in complex geometries}, Annual research briefs, Center for Turbulence
  Research (1997).

\bibitem{FADLUN200035}
E.~Fadlun, R.~Verzicco, P.~Orlandi, J.~Mohd-Yusof,
  \href{http://www.sciencedirect.com/science/article/pii/S0021999100964842}{Combined
  immersed-boundary finite-difference methods for three-dimensional complex
  flow simulations}, Journal of Computational Physics 161~(1) (2000) 35 -- 60.
\newblock \href {http://dx.doi.org/http://dx.doi.org/10.1006/jcph.2000.6484}
  {\path{doi:http://dx.doi.org/10.1006/jcph.2000.6484}}.
\newline\urlprefix\url{http://www.sciencedirect.com/science/article/pii/S0021999100964842}

\bibitem{Majumdar2001}
S.~Majumdar, G.~Iaccarino, P.~Durbin,
  \href{http://ctr.stanford.edu/ResBriefs01/sekhar.pdf}{{RANS solvers with
  adaptive structured boundary non-conforming grids}}, Annual Research Briefs
  (2001) 353--366.
\newline\urlprefix\url{http://ctr.stanford.edu/ResBriefs01/sekhar.pdf}

\bibitem{Tseng_ghostCell}
Y.-H. Tseng, J.~H. Ferziger,
  \href{http://dx.doi.org/10.1016/j.jcp.2003.07.024}{A ghost-cell immersed
  boundary method for flow in complex geometry}, J. Comput. Phys. 192~(2)
  (2003) 593--623.
\newblock \href {http://dx.doi.org/10.1016/j.jcp.2003.07.024}
  {\path{doi:10.1016/j.jcp.2003.07.024}}.
\newline\urlprefix\url{http://dx.doi.org/10.1016/j.jcp.2003.07.024}

\bibitem{YE1999209}
T.~Ye, R.~Mittal, H.~Udaykumar, W.~Shyy,
  \href{http://www.sciencedirect.com/science/article/pii/S0021999199963568}{An
  accurate cartesian grid method for viscous incompressible flows with complex
  immersed boundaries}, Journal of Computational Physics 156~(2) (1999) 209 --
  240.
\newblock \href {http://dx.doi.org/http://dx.doi.org/10.1006/jcph.1999.6356}
  {\path{doi:http://dx.doi.org/10.1006/jcph.1999.6356}}.
\newline\urlprefix\url{http://www.sciencedirect.com/science/article/pii/S0021999199963568}

\bibitem{Hartmann2011}
D.~Hartmann, M.~Meinke, W.~Schröder, A strictly conservative cartesian
  cut-cell method for compressible viscous flows on adaptive grids, Computer
  Methods in Applied Mechanics and Engineering 200~(9-12) (2011) 1038--1052.
\newblock \href {http://dx.doi.org/10.1016/j.cma.2010.05.015}
  {\path{doi:10.1016/j.cma.2010.05.015}}.

\bibitem{SCHNEIDERS2013786}
L.~Schneiders, D.~Hartmann, M.~Meinke, W.~Schröder,
  \href{http://www.sciencedirect.com/science/article/pii/S0021999112005839}{An
  accurate moving boundary formulation in cut-cell methods}, Journal of
  Computational Physics 235 (2013) 786 -- 809.
\newblock \href {http://dx.doi.org/http://dx.doi.org/10.1016/j.jcp.2012.09.038}
  {\path{doi:http://dx.doi.org/10.1016/j.jcp.2012.09.038}}.
\newline\urlprefix\url{http://www.sciencedirect.com/science/article/pii/S0021999112005839}

\bibitem{TANG2003567}
H.~Tang, S.~C. Jones, F.~Sotiropoulos,
  \href{http://www.sciencedirect.com/science/article/pii/S0021999103003310}{An
  overset-grid method for 3d unsteady incompressible flows}, Journal of
  Computational Physics 191~(2) (2003) 567 -- 600.
\newblock \href
  {http://dx.doi.org/http://dx.doi.org/10.1016/S0021-9991(03)00331-0}
  {\path{doi:http://dx.doi.org/10.1016/S0021-9991(03)00331-0}}.
\newline\urlprefix\url{http://www.sciencedirect.com/science/article/pii/S0021999103003310}

\bibitem{overset1}
K.~Nakahashi, F.~Togashi, D.~Sharov, Intergrid-boundary definition method for
  overset unstructured grid approach, AIAA Journal 38~(11) (2000) 2077 -- 2084.
\newblock \href {http://dx.doi.org/https://doi.org/10.2514/2.869}
  {\path{doi:https://doi.org/10.2514/2.869}}.

\bibitem{overset2}
R.~Meakin, Moving body overset grid methods for complete aircraft tiltrotor
  simulations, 11th Computational Fluid Dynamics Conference, Fluid Dynamics and
  Co-located Conferences\href
  {http://dx.doi.org/https://doi.org/10.2514/6.1993-3350}
  {\path{doi:https://doi.org/10.2514/6.1993-3350}}.

\bibitem{Wesseling:2000:PCF:1211014}
P.~Wesseling, Principles of Computational Fluid Dynamics, Springer-Verlag New
  York, Inc., Secaucus, NJ, USA, 2000.

\bibitem{ferziger2001}
J.~Ferziger, M.~Peric,
  \href{https://books.google.co.uk/books?id=1D3EQgAACAAJ}{Computational Methods
  for Fluid Dynamics}, Springer Berlin Heidelberg, 2001.
\newline\urlprefix\url{https://books.google.co.uk/books?id=1D3EQgAACAAJ}

\bibitem{Hirt1974}
C.~W. {Hirt}, A.~A. {Amsden}, J.~L. {Cook}, {An Arbitrary Lagrangian-Eulerian
  Computing Method for All Flow Speeds}, Journal of Computational Physics 14
  (1974) 227--253.
\newblock \href {http://dx.doi.org/10.1016/0021-9991(74)90051-5}
  {\path{doi:10.1016/0021-9991(74)90051-5}}.

\bibitem{ALEbook}
J.~Donea, A.~Huerta, J.-P. Ponthot, A.~Rodríguez-Ferran,
  \href{http://dx.doi.org/10.1002/0470091355.ecm009}{Arbitrary
  Lagrangian-Eulerian Methods}, John Wiley \& Sons, Ltd, 2004.
\newblock \href {http://dx.doi.org/10.1002/0470091355.ecm009}
  {\path{doi:10.1002/0470091355.ecm009}}.
\newline\urlprefix\url{http://dx.doi.org/10.1002/0470091355.ecm009}

\bibitem{patankar1980numerical}
S.~Patankar, \href{https://books.google.fr/books?id=5JMYZMX3OVcC}{Numerical
  Heat Transfer and Fluid Flow}, Series in computational methods in mechanics
  and thermal sciences, Taylor \& Francis, 1980.
\newline\urlprefix\url{https://books.google.fr/books?id=5JMYZMX3OVcC}

\bibitem{IssaPISO}
R.~I. Issa, \href{http://dx.doi.org/10.1016/0021-9991(86)90099-9}{Solution of
  the implicitly discretised fluid flow equations by operator-splitting}, J.
  Comput. Phys. 62~(1) (1986) 40--65.
\newblock \href {http://dx.doi.org/10.1016/0021-9991(86)90099-9}
  {\path{doi:10.1016/0021-9991(86)90099-9}}.
\newline\urlprefix\url{http://dx.doi.org/10.1016/0021-9991(86)90099-9}

\bibitem{valve}
FMCTechnologies,
  \href{http://www.arm-tex.com/assets/pdfs/FMC/Weco-Butterfly-Valves-General-Catalog.pdf}{Weco\textregistered
  butterfly valve} (2017).
\newline\urlprefix\url{http://www.arm-tex.com/assets/pdfs/FMC/Weco-Butterfly-Valves-General-Catalog.pdf}

\bibitem{aboutFlow}
Adjoint based optimization of industrial unsteady flows.
  \underline{http://aboutflow.sems.qmul.ac.uk}.

\end{thebibliography}
